\documentclass[useAMS,usenatbib]{mn2e}

\newcommand\article{{\it Article}}
\newcommand\ges{{\it Gaia-ESO Survey}}
\newcommand\corot{{\it CoRoT}}
\newcommand\kepler{{\it Kepler}}
\newcommand\nd{$\cdots$}

\usepackage{amssymb}
\usepackage{graphicx}
\usepackage{caption}
\usepackage{subcaption}

\title[Revisiting the Li-rich giant problem]{The Gaia-ESO Survey: Revisiting the Li-rich giant problem}
\author[Casey et al.]{A.~R. Casey$^1$\thanks{E-mail: arc@ast.cam.ac.uk},
G. Ruchti$^2$, T. Masseron$^1$, S. Randich$^3$, G. Gilmore$^1$, K. Lind$^{4,5}$,\newauthor
G.~M. Kennedy$^1$, S.~E. Koposov$^{1}$, A. Hourihane$^1$, E. Franciosini$^3$, J.~R. Lewis$^{1}$,\newauthor
L. Magrini$^{3}$, L. Morbidelli$^3$, G.~G. Sacco$^3$, C.~C. Worley$^1$, S. Feltzing$^2$, \newauthor
R.~D. Jeffries$^{6}$, A. Vallenari$^{7}$, T. Bensby$^2$,  A. Bragaglia$^{8}$, E. Flaccomio$^{9}$, \newauthor
P.~Francois$^{10}$, A.~J. Korn$^5$, A. Lanzafame$^{11}$, E. Pancino$^{3,12}$, A. Recio-Blanco$^{13}$,\newauthor
R. Smiljanic$^{14}$, G. Carraro$^{15}$, M.~T. Costado$^{16}$, F. Damiani$^{9}$, P. Donati$^{8,17}$, \newauthor
A. Frasca$^{18}$, P. Jofr\'e$^1$, C. Lardo$^{19}$, P. de Laverny$^{13}$, L. Monaco$^{20}$, L. Prisinzano$^{9}$,\newauthor
L. Sbordone$^{21,22}$, S.~G. Sousa$^{23}$, G. Tautvai\v{s}ien\.{e}$^{24}$, S. Zaggia$^{7}$, T. Zwitter$^{25}$, \newauthor
E. Delgado Mena$^{23}$, Y. Chorniy$^{24}$, S.~L. Martell$^{26}$, V. Silva Aguirre$^{27}$,  A. Miglio$^{28}$, \newauthor
C. Chiappini$^{29}$, J. Montalban$^{30}$, T. Morel$^{31}$, M. Valentini$^{29}$
\\
Affiliations can be found at the end of this Article.
}
\begin{document}

\def\aj{AJ}                   
\def\araa{ARA\&A}             
\def\apj{ApJ}                 
\def\apjl{ApJ}                
\def\apjs{ApJS}               
\def\ao{Appl.Optics}          
\def\apss{Ap\&SS}             
\def\aap{A\&A}                
\def\aapr{A\&A~Rev.}          
\def\aaps{A\&AS}              
\def\azh{AZh}                 
\def\baas{BAAS}               
\def\jrasc{JRASC}             
\def\memras{MmRAS}            
\def\mnras{MNRAS}             
\def\nat{Nature}
\def\procspie{Proceedings of SPIE}
\def\pra{Phys.Rev.A}          
\def\prb{Phys.Rev.B}          
\def\prc{Phys.Rev.C}          
\def\prd{Phys.Rev.D}          
\def\prl{Phys.Rev.Lett}       
\def\pasp{PASP}               
\def\pasj{PASJ}               
\def\qjras{QJRAS}             
\def\skytel{S\&T}             
\def\solphys{Solar~Phys.}     
\def\sovast{Soviet~Ast.}      
\def\ssr{Space~Sci.Rev.}      
\def\zap{ZAp}                 
\let\astap=\aap
\let\apjlett=\apjl
\let\apjsupp=\apjs
\def\aj{AJ}                   
\def\araa{ARA\&A}             
\def\apj{ApJ}                 
\def\apjl{ApJ}                
\def\apjs{ApJS}               
\def\ao{Appl.Optics}          
\def\apss{Ap\&SS}             
\def\aap{A\&A}                
\def\aapr{A\&A~Rev.}          
\def\aaps{A\&AS}              
\def\azh{AZh}                 
\def\baas{BAAS}               
\def\jrasc{JRASC}             
\def\memras{MmRAS}            
\def\mnras{MNRAS}            
\def\pasa{PASA}  
\def\pra{Phys.Rev.A}          
\def\prb{Phys.Rev.B}          
\def\prc{Phys.Rev.C}          
\def\prd{Phys.Rev.D}          
\def\prl{Phys.Rev.Lett}       
\def\pasp{PASP}               
\def\pasj{PASJ}               
\def\qjras{QJRAS}             
\def\skytel{S\&T}             
\def\solphys{Solar~Phys.}     
\def\sovast{Soviet~Ast.}      
\def\ssr{Space~Sci.Rev.}      
\def\zap{ZAp}                 
\let\astap=\aap
\let\apjlett=\apjl
\let\apjsupp=\apjs

\date{Accepted 2016 XX XX. Received 2016 February XX; in original form 2016 May XX}

\pagerange{\pageref{firstpage}--\pageref{lastpage}} \pubyear{2015}

\maketitle

\label{firstpage}

\begin{abstract}
The discovery of lithium-rich giants contradicts expectations from canonical
stellar evolution. Here we report on the serendipitous discovery of 20 Li-rich 
giants observed during the \ges, which includes the first nine Li-rich giant
stars known towards the \corot\ fields.  Most of our Li-rich giants have 
near-solar metallicities, and stellar parameters consistent with being before
the luminosity bump. This is difficult to reconcile with deep mixing models 
proposed to explain lithium enrichment, because these models can only operate 
at later evolutionary stages: at or past the luminosity bump.  In an effort to 
shed light on the Li-rich phenomenon, we highlight recent evidence of the tidal 
destruction of close-in hot Jupiters at the sub-giant phase.  We note that 
when coupled with models of planet accretion, the observed destruction of hot
Jupiters actually predicts the existence of Li-rich giant stars, and suggests
Li-rich stars should be found early on the giant branch and occur more 
frequently with increasing metallicity. A comprehensive review of all known 
Li-rich giant stars reveals that this scenario is consistent with the data. 
However more evolved or metal-poor stars are less likely to host close-in 
giant planets, implying that their Li-rich origin requires an alternative 
explanation, likely related to mixing scenarios rather than external phenomena.
\end{abstract}


\section{Introduction}
\label{sec:introduction}

Lithium is a fragile element which cannot be easily replenished.  Given its
fragility, canonical stellar evolution models predict that a star's Li 
abundance should decrease as it ascends the giant branch.  Observations since
\citet{Bonsack_1959} have repeatedly confirmed these predictions.  However, 
Population~I stars show lithium abundances approximately ten times higher than
older Population~II stars, implying some kind of Galactic lithium enrichment.
More puzzlingly, there exists a growing number of giant stars with lithium 
abundances that are near to, or exceed Big Bang nucleosynthesis predictions. 
Although rare, these stars constitute a fundamental outstanding problem for 
stellar evolution.

Stellar evolution theory suggests that the depth of the convective envelope 
increases when a star leaves the main-sequence.  In doing so, the star 
experiences first dredge-up: material from deep internal layers is mixed
towards the surface \citep{Iben_1967a,Iben_1967b}. The inner material is
hot enough that Li has been destroyed, therefore first dredge up dilutes the surface Li
abundance.  Consequently, stellar evolution theory predicts the observable 
Li abundance should be $\sim$1.5~dex lower for evolved stars than their 
main-sequence counterparts. \citep[e.g.,][]{Iben_1967a,Lagarde_2012}
Stars on the upper red giant branch (RGB) may be even more depleted in Li
due to mixing occurring just after the RGB bump \citep{Sweigart_Mengel_1979,
Charbonnel_1994,Charbonnel_1995}.  Other changes to surface abundances are 
also predicted: increases in $^4$He, $^{14}$N, $^{13}$C, and decreases in
$^{12}$C \citep{Iben_1964,Chaname_2005,Charbonnel_2006,Charbonnel_Lagarde_2010,
Karakas_2010,Lattanzio_2015}.  Detailed observations have repeatedly provided
convincing evidence of these predictions \citep[e.g.,][]{Lambert_1980,
Spite_1982,Gratton_2000,Lind_2009b,Mucciarelli_2012,Tautvaisiene_2013}.

The existence of Li-rich (A(Li) $\gtrsim 2$) giant stars
implies an additional mechanism that produces and/or preserves surface Li.
This process may be internal or external.  In the right
conditions stars can produce Li internally through the Cameron-Fowler mechanism
\citep{Cameron_Fowler_1971}: $^3{\rm He}(\alpha,\gamma)^7{\rm Be}(e^-,\nu)^7{\rm Li}$.
The temperature must be hot enough for $^7$Be to be produced, but $^7$Be must
be quickly transported towards cooler regions so that fresh $^7$Li can be 
created without being immediately destroyed by proton capture.  The 
Cameron-Fowler mechanism can operate in red giants in two different 
stages.  During hot bottom burning (HBB), the bottom of the convective envelope 
is hot enough for $^7$Be production.  The convection carries $^7$Be
to cooler regions where it can capture an electron to produce $^7$Li.
In the absence of HBB, a radiative zone exists between the
shell and the convective envelope.  A mechanism is then required to mix material
down to the outer part of the shell -- where temperatures are high enough to produce
$^7$Be -- and then fresh $^7$Be must be mixed across the radiative 
zone to the convective envelope.  The mechanism for mixing through the 
radiative zone is under debate, but various mechanisms are collectively referred to as `deep-mixing'
or `extra-mixing'.  Moreover, because the conditions required to produce $^7$Li
are also sufficient to destroy it (e.g., by mixing fresh $^7$Li back to hotter
regions), the level of subsequent Li-enhancement due to extra mixing is 
critically sensitive to the mixing speed, geometry, and episodicity 
\citep[e.g.,][]{Sackmann_Boothroyd_1999}.

Several scenarios have been proposed to reconcile the existence of Li-rich giant
stars, including ones that aim to minimise the amount of partial burning 
(i.e., preserve existing Li).  However using \textit{Hipparcos} parallaxes
\citep{van_Leeuwen_2007} and stellar tracks to precisely estimate stellar masses
and evolutionary states, \citet{Charbonnel_2000} highlight fifteen Li-rich giants
where Li preservation is insufficient: a Li-\emph{production} mechanism is required
to match the data.  While precise, Li abundance measurements can be limited by
the stellar tracks employed (e.g., including the horizontal or asymptotic branch
for low-mass stars), emphasising the need for accurate knowledge of the 
evolutionary status. \citet{Charbonnel_2000} propose two distinct episodes of
Li production that depend on the stellar mass.  For low-mass RGB stars at the
bump in the luminosity function, the outward-moving hydrogen shell burns through
the mean molecular weight discontinuity produced during first dredge up, enabling
extra mixing and facilitating the Cameron-Fowler mechanism. However in 
intermediate-mass stars, the composition discontinuity is not destroyed 
until after the star begins core He burning.  For this reason extra mixing
can only be induced in intermediate-mass asymptotic giant branch (AGB) stars when 
the convective envelope deepens at the base of the AGB.  While these scenarios
explain the necessary internal conditions required to produce and transport Li
to the photosphere, they do not speculate on the actual mechanism that drives 
the mixing \citep[however see][]{Charbonnel_Zahn_2007}.  \citet{Palacios_2006}
have shown that rotation alone is insufficient to produce the observed Li
abundances, implying that an additional mechanism is required to induce the
extra mixing.

Thermohaline mixing has been proposed as a mechanism to drive extra mixing
at the bump in the giant branch luminosity function. In addition to removing
any existing molecular weight gradient, an inversion in the molecular weight 
gradient is produced, which drives thermohaline mixing \citep{Eggleton_2006}.
In contrast, \citet{Denissenkov_2003} incorporate 
diffusion and shear-driven mixing to facilitate extra-mixing in low-mass RGB 
stars.  Their prescription relies on main-sequence stars (as the precursors of 
upper RGB stars) to possess rapidly rotating radiative cores.  Instead of 
encouraging interactions between different mass shells 
\citep[e.g.,][]{Charbonnel_2000}, \citet{Denissenkov_2003} require the 
specific angular momentum to be conserved in each shell during the star's 
evolution.  This situation would therefore permit a reservoir of angular 
momentum which could later induce deep mixing.

\citet{Palacios_2001} proposed that internal instabilities occurring near the
luminosity bump were sufficient to produce additional Li. Specifically,
internally-produced $^7$Be could be transported to a nearby convective region
where $^7$Li is produced, but immediately destroyed by proton capture.  In effect,
a thin burning layer of Li is created, where $^7{\rm Li}(p,\alpha)\alpha$
becomes the dominant reaction, increasing the local temperature and the level of
meridional circulation.  The molecular weight gradient is eventually destroyed,
allowing for deep mixing to occur.  While promising, this scenario requires
an arbitrary and substantially large change in diffusion rates.  A significant
amount of mass-loss is expected as a consequence of this scenario, as well as
an excess in infrared colours.  Given extensive investigations into the (lack of)
association between far infrared excesses and Li-rich giants, it would appear
this scenario may be unlikely, unless the infrared excess phase is short 
\citep{Rebull_2015,de_la_Reza_2015}.

\begin{figure*}
\includegraphics[width=\textwidth]{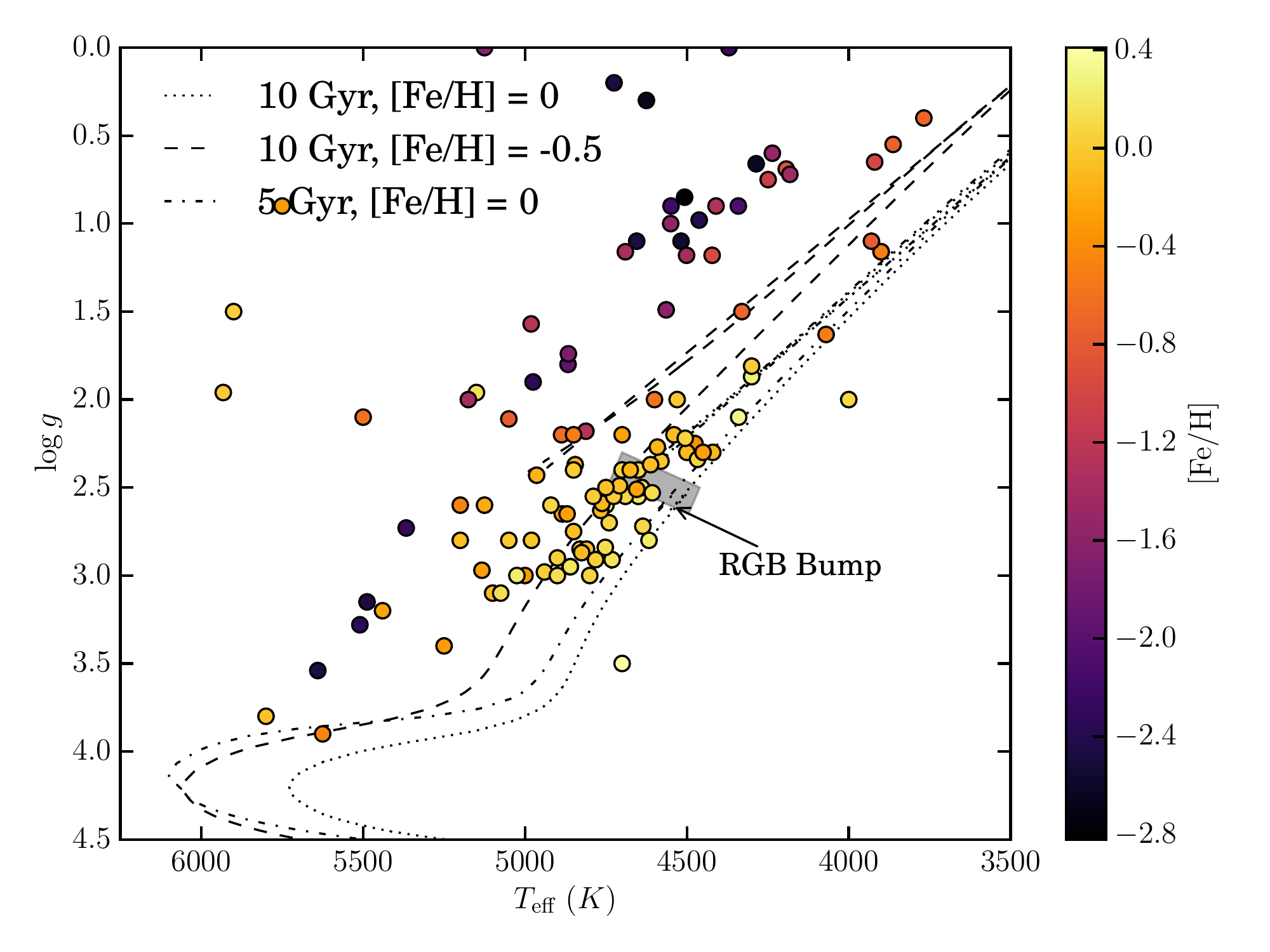}
\caption{Stellar parameters ($T_{\rm eff}$, $\log{g}$, [Fe/H]) for all 127 giant stars in the literature with ${\rm A(Li)} > 2$ (in LTE or non-LTE). Discoveries in this study are included. Two 10~Gyr PARSEC isochrones \citep[][assuming $Z$ is directly proportional to Fe and the PARSEC default $Y = 0.2485+1.78Z$]{Bressan_2012} of different metallicity are also shown. Subject to selection functions, this figure indicates that Li-rich giant stars occur more frequently before the luminosity bump on the giant branch, and at near-solar metallicities.}
\label{fig:literature}

\end{figure*}

The extra mixing required may be induced by external phenomena.  The ingestion 
of a massive planet or brown dwarf would contribute significant angular momentum
to the system, producing additional Li before it is destroyed by convection
\citep{Alexander_1967,Siess_Livio_1999a,Siess_Livio_1999b,Denissenkov_Weiss_2000,
Denissenkov_Herwig_2004,Carlberg_2010}.  In this scenario the planet is assumed
to be dissipated at the base of the convective envelope of a giant star, causing
the star to substantially expand in size.  If the accretion rate is high, HBB
can be triggered.  The predicted observational signatures vary depending on the 
accretion rate and the ingestion angle of the planet/dwarf.  However, the
predicted observables include increased mass loss and/or the ejection of a shell
(and therefore a subsequent phase of infrared emission), an increase in the 
$^7$Li surface abundance, potential stellar metallicity enrichment, possibly 
increased rotational velocity due to the transfer of angular momentum, and less 
discernible effects such as the generation of magnetic fields \citep[however 
see][]{Lebre_2009} or changes to the morphology of the horizontal branch.  
\citet{Siess_Livio_1999a,Siess_Livio_1999b} argue that the planet/dwarf star 
accretion scenario is not limited to a single evolutionary stage, allowing for
Li-rich giants to exist on the red giant and the asymptotic giant branch.  It 
can also advantageously explain stars with either high or low rotational 
velocities, depending on the extent that magnetic braking has influenced 
spin-down.  However, there has been no discussion in the literature on how this 
scenario alone relates to why Li-rich giants tend to appear more frequently just
below the RGB bump (e.g., see Figure \ref{fig:literature}). Similarly, there has
been no discussion of links between Li-rich giant stars and the properties or
occurrence rates of exoplanet host stars.

\citet{Martin_1994} propose a novel external mechanism to reconcile observations
of Li-rich giant stars.  High Li abundances detected in the secondaries of a 
stellar-mass black hole \citep{Martin_1992} and neutron star \citep{Martin_1994}
candidates led to the postulation that Li could be produced during a supernova 
explosion \citep[see also][]{Tajitsu_2015}, or through $\alpha$-$\alpha$ 
reactions during strong outbursts from a transient X-ray binary system.  These 
conditions could be sufficiently energetic to induce cosmic-ray spallation and
produce Li \citep{Walker_1985}.  Li would presumably be accreted to the edge of
the convective envelope of the secondary thereby producing a Li-rich giant star, 
potentially at any stage across the RGB, with low rotational velocities.  
A consequence of Li spallation is that beryllium and boron would also be
created.  To date, no Li-rich giant star has been found to have Be enhancement
\citep{De_Medeiros_1997,Castilho_1999,Melo_2005}.  Finally, although no 
long-term radial velocity studies have been conducted, the non-detection of
a white dwarf companion in the vicinity of a present-day Li-rich giant star 
weakens this idea.

Observations have been key to guiding models that can explain Li-rich giant
stars.  Unfortunately most Li-rich giant stars are not distinguishable by their photometric colours, 
therefore they cannot be efficiently selected solely on the basis of photometry.
Early observations of far-infrared colours showed that many Li-rich stars 
show far-infrared excesses \citep{de_La_Reza_1996,de_La_Reza_1997},
suggesting that the Li-rich phase was associated with a mass-loss event.  
However later K-giant selections based on far-infrared colour excesses did not
reveal any new Li-rich stars \citep{Fekel_Watson_1998,Jasniewicz_1999}.  
\citet{Rebull_2015} studied this phenomenon extensively and revealed that the
largest infrared excesses do indeed appear in Li-rich K giants \citep[typically 
with fast rotation, see also][]{Fekel_Balachandran_1993}, although very few 
Li-rich K giants show any infrared excess.  \citet{Kumar_2015} came to the same
conclusion from a study of $\sim 2000$ K giants. Therefore, if mass-loss or dust 
shell production is a regular consequence of the Li-enrichment mechanism, the 
infrared excess phase must be short \citep{de_la_Reza_2015}.

Discoveries of Li-rich giant stars have been slow relative to advances in 
modelling.  Their sparsity is partly to blame: only 1\% of slow rotating K giant
stars are Li-rich \citep[although $\sim$50\% of rapid rotating K giants are
Li-rich, see][]{Drake_2002,Lebre_2006}.  For this reason most discoveries have been 
reported individually, although they cover all major components of the Galaxy: 
towards the bulge \citep{McWilliam_1994,Uttenthaler_2007,Gonzalez_2009}, disk 
\citep{Monaco_2011}, as well as plenty in the field\footnote{For example, see 
\citet{Wallerstein_Sneden_1982,Luck_1982,Hanni_1984,Andrievsky_1999,
Balachandran_2000,Reyniers_Van_Winckel_2001,Lebre_2009,Alcala_2011,Kumar_2011,
Ruchti_2011,Kovari_2013,Liu_2014,Adamow_2015}.}.  Li-rich giant stars have also
been found in dwarf galaxies \citep{Kirby_2016}, where the most metal-poor 
(${\rm [Fe/H]} \approx -2.8$) Li-rich giant star known has been found 
\citep{Kirby_2012}.  Interestingly, despite large observational programs dedicated 
to obtaining high-quality spectra in clusters, fewer than ten Li-rich giants have 
been discovered in globular clusters \citep[2 in NGC~362, M3-IV101, M5-V42, M68-A96, etc;][]{Smith_1999,DOrazi_2015,
Kraft_1999,Carney_1998,Ruchti_2011,Kirby_2016}, and just five in open clusters
\citep[NGC~7789-K301, Berkeley~21, M~67, Trumpler~5, and NGC~6819;][respectively]{
Pilachowski_1986,Hill_1999,Canto_Martins_2006,Monaco_2014,Twarog_2013,
Carlberg_2015}\footnote{See also \citet{Delgado_Mena_2015}.}.

Because the mixing mechanisms required to produce Li-rich giant stars are sensitive
to the evolutionary stage, asteroseismology is a promising field to distinguish
proposed mixing scenarios.  To date five Li-rich giant stars have been 
discovered in the \kepler\ field \citep{Martell_2013,Silva_Aguirre_2014,
Jofre_2015,Twarog_2013,Carlberg_2015}.  However only two have benefited from
seismic information. One Li-rich giant star has been shown to host a He-burning
core, suggesting that Li production may have occurred through non-canonical 
mixing at the RGB tip \citep{Kumar_2011}, possibly during the helium flash 
\citep[see also][]{Cassisi_2016}. In contrast, seismic data for the Li-rich 
star KIC~9821622 has shown that it does \emph{not} host a He-burning core, and sits
just before the luminosity bump on the giant branch \citep{Jofre_2015}.  Clearly, 
a larger sample of Li-rich giant stars with detectable solar-like oscillations is needed.

Large scale spectroscopic surveys are ideal vehicles for increasing the sample of 
known Li-rich giant stars.  In this \article\ we report the serendipitous discovery 
of 20 previously unknown Li-rich giants in the \ges. Four were observed with the UVES
spectrograph, and the remainder using GIRAFFE. This constitutes one of the largest 
sample of Li-rich giant stars ever discovered. This \article\ is organised in the 
following manner.  In Section \ref{sec:data} we describe the data and analysis.  
We discuss the evolutionary stage and associated environments for all stars in our 
sample in Section \ref{sec:discussion}, before commenting on the likelihood of 
different Li production mechanisms. We conclude in Section \ref{sec:conclusions}.

\section{Data \& Analysis}
\label{sec:data}

The \ges\ \citep[][ESO programs 188.B-3002 and 193.B-0936]{Gilmore_2012,
Randich_2013} is a $\sim$300-night program that simultaneously uses the UVES and
GIRAFFE spectrographs \citep{Dekker_2000,Pasquini_2000} on the Very Large 
Telescope in Chile to obtain high-resolution optical spectra for $>$100,000 
stars in the Galaxy.  Targets from all stellar populations are observed.

We searched the fourth internal data release (iDR4) of the \ges\ for giant stars
with peculiarly high lithium abundances.  We restricted our search to K-type 
giant stars with Li measurements (i.e., not upper limits) where 
A(Li LTE) $\gtrsim 2$.  Our search revealed 4 bonafide 
Li-rich giant stars observed with UVES, and 16 observed with GIRAFFE.  A 
cross-match of the \textit{Survey} observing logs reveals these spectra were 
obtained in good seeing (0.6-0.9\arcsec) throughout 2013-2014.  Standard data 
reduction procedures were employed, as detailed in \citet{Sacco_2014} and 
\citet{Lewis_2016}.  The S/N of the spectra range from $\approx$30 to 
$\approx$100.

The \ges\ employs multiple analysis pipelines to produce a robust ensemble
measurement of the stellar parameters (Table \ref{tab:stellar-parameters}: 
$T_{\rm eff}$, $\log{g}$, ${\rm [M/H]}$) and detailed chemical abundances.
The analysis of FGK-type stars within the \textit{Survey} is split between
different working groups (WGs): WG10 analyses FGK-type stars observed with 
GIRAFFE, WG11 analyses FGK-type stars observed with UVES 
\citep{Smiljanic_2014}, and WG12 analyses pre-main-sequence candidates
\citep{Lanzafame_2015} -- irrespective of whether they were observed with
GIRAFFE or UVES.  Within each WG there are multiple analysis nodes.  A node 
consists of a sufficiently distinct pipeline, and expert spectroscopists
that are familiar with the pipeline employed.  All nodes provide estimates 
of the stellar parameters and/or detailed chemical abundances.  For the 
\ges\ iDR4, there are up to six nodes for WG10, and eleven for WG11.

There are some commonalities between the nodes.  The MARCS 1D model 
atmospheres \citep{Gustafsson_2008} are used by all nodes, the same atomic line data 
\citep{Ruffoni_2014,Heiter_2015} and solar abundances \citep{Grevesse_2007} 
are employed, and where relevant, the same grid of synthetic spectra is used.
The WG10/GIRAFFE nodes are provided initial guesses of the stellar parameters 
from a pre-processing pipeline.  The data reduction procedure also produces 
normalised spectra for all nodes, however some nodes opted to repeat or 
redo the normalisation.

The spectral analysis is performed in two consecutive stages.  The stellar 
parameters reported by each node are homogenised to produce an ensemble
measurement of stellar parameters for a given star.  Those homogenised 
measurements are then returned to the nodes, at which point the detailed chemical 
abundances are calculated using the homogenised stellar parameters.  Appropriate
data are accounted for during the abundance determination of each line or 
element (e.g., hyperfine structure, the Fe 6707.4~\AA\ blend for Li abundances, 
etc).  Individual abundances are subsequently homogenised, producing a single 
set of abundance measurements for all co-investigators of the \textit{Survey} to
use.  In both stages (stellar parameters, chemical abundances), the 
homogenisation procedure identifies erroneous node measurements, accounts for 
the covariance between sources of measurements, and quantifies or minimises 
systematics present in the data.  Most critically, the top-level homogenisation 
(performed by WG15) ensures that results from multiple WGs are on a consistent, 
comparable scale.  Details of the analysis nodes, work structure and
homogenisation procedure for the previous WG11 data release is presented in 
\citet{Smiljanic_2014}.  A full description of the homogenisation procedure for 
UVES iDR4 data will be presented in \citet{Casey_2016}.

\subsection{Characterisation and Evolutionary Status of Li-Rich Stars}

Our sample of bonafide Li-rich giant stars includes targets analysed by WG10,
WG11, and WG12.  While the WG12 group include experts on the analysis of
pre-main-sequence stars, they are also specialists in standard FGK-type star
analyses.  This is important to note, as not all stars targeted by WG12 are
later found to be pre-main-sequence stars; some stars targeted by WG12 are
actually standard FGK-type stars.  Half (10) of our Li-rich giant stars
were analysed by WG10 or WG11. The remainder were targeted as pre-main-sequence 
candidates towards young clusters, but were later found to be giant stars 
that are likely non-members of those clusters (see below).  Their evolved 
nature is indicative from their stellar parameters, the empirical $\gamma$-index 
\citep[we required $\gamma > 1.01$;][]{Damiani_2014}, and lack of H-$\alpha$ 
emission (a youth indicator for pre-main-sequence stars).

\begin{figure*}
\includegraphics[width=\textwidth]{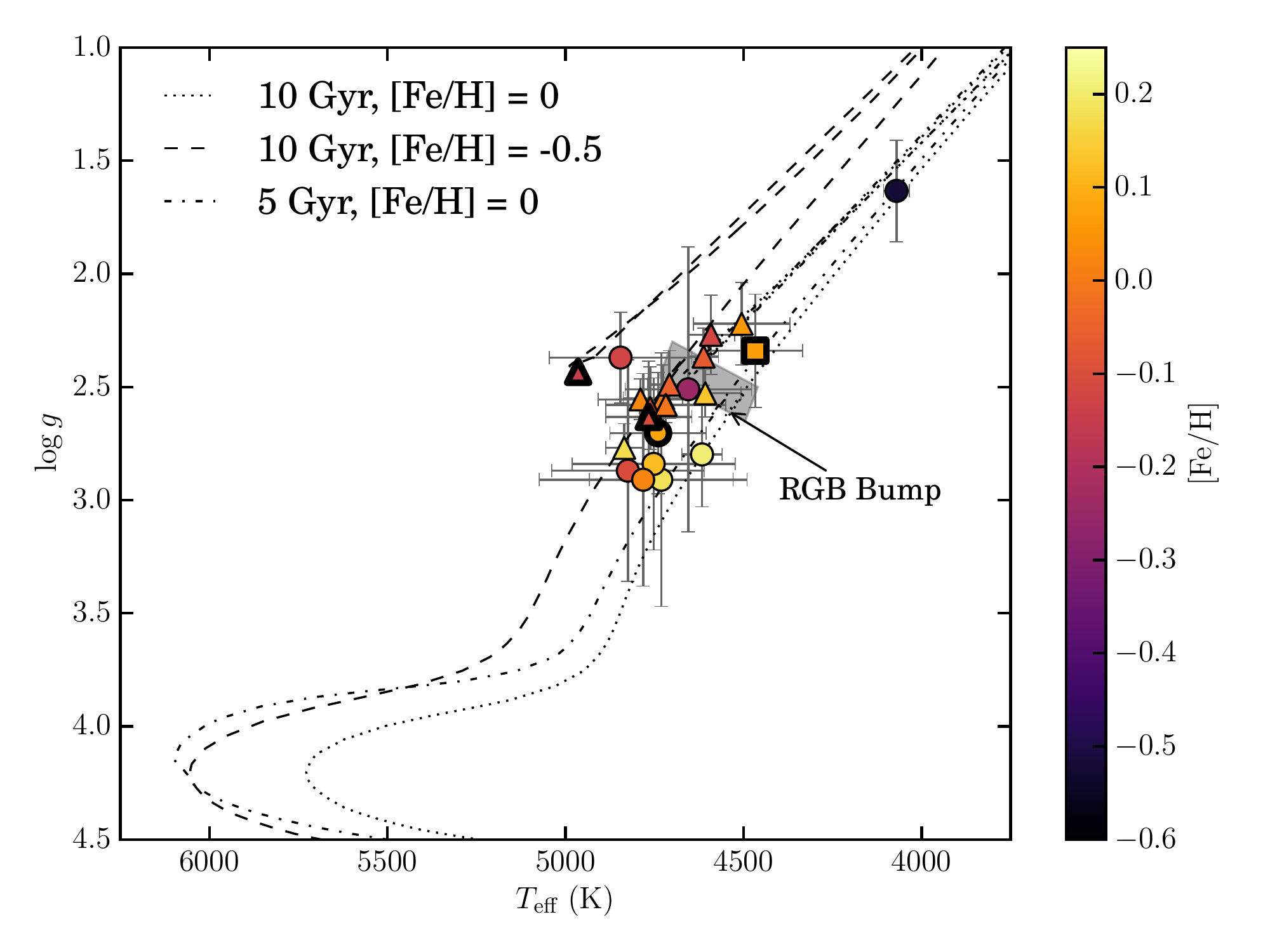}
\caption{Stellar parameters ($T_{\rm eff}$, $\log{g}$, [Fe/H]) for all Li-rich giant stars in our sample, shown upon 5~Gyr and 10~Gyr PARSEC isochrones with $Y=0.2485+1.78Z$ \citep{Bressan_2012}. We highlight the approximate
location of the RGB bump from the isochrones shown. Markers are coloured by their metallicity. Thick edges indicate the star was observed with UVES. The bulge Li-rich star is indicated by a square marker,  and circular markers indicate \corot\ targets.}
\label{fig:stellar-params}

\end{figure*}

Most stars in our sample lie below the RGB bump (Figure \ref{fig:stellar-params}),
consistent with previous studies of Li-rich giant stars with near-solar 
metallicities (Figure \ref{fig:literature}).  Some stars are exceptions: 
18033785--3009201 was observed with UVES and lies just above the RGB bump, 
near the clump.  19230935+0123293 has a similar surface gravity, but is hotter
and more consistent with being a red clump (RC) or AGB star.  19301883--0004175 
is the coolest and most metal-poor (${\rm [Fe/H]} = -0.52$) Li-rich giant star
in our sample.  Our stellar parameters place 19301883--0004175 slightly red-ward
(below) of the isochrone.  Given this star is in the \corot\ field, combining 
asteroseismic oscillations with the high-quality \ges\ spectra would be 
advantageous to firmly establish the evolutionary state of this highly evolved
Li-rich giant star.

\begin{figure*}
\includegraphics[width=\textwidth]{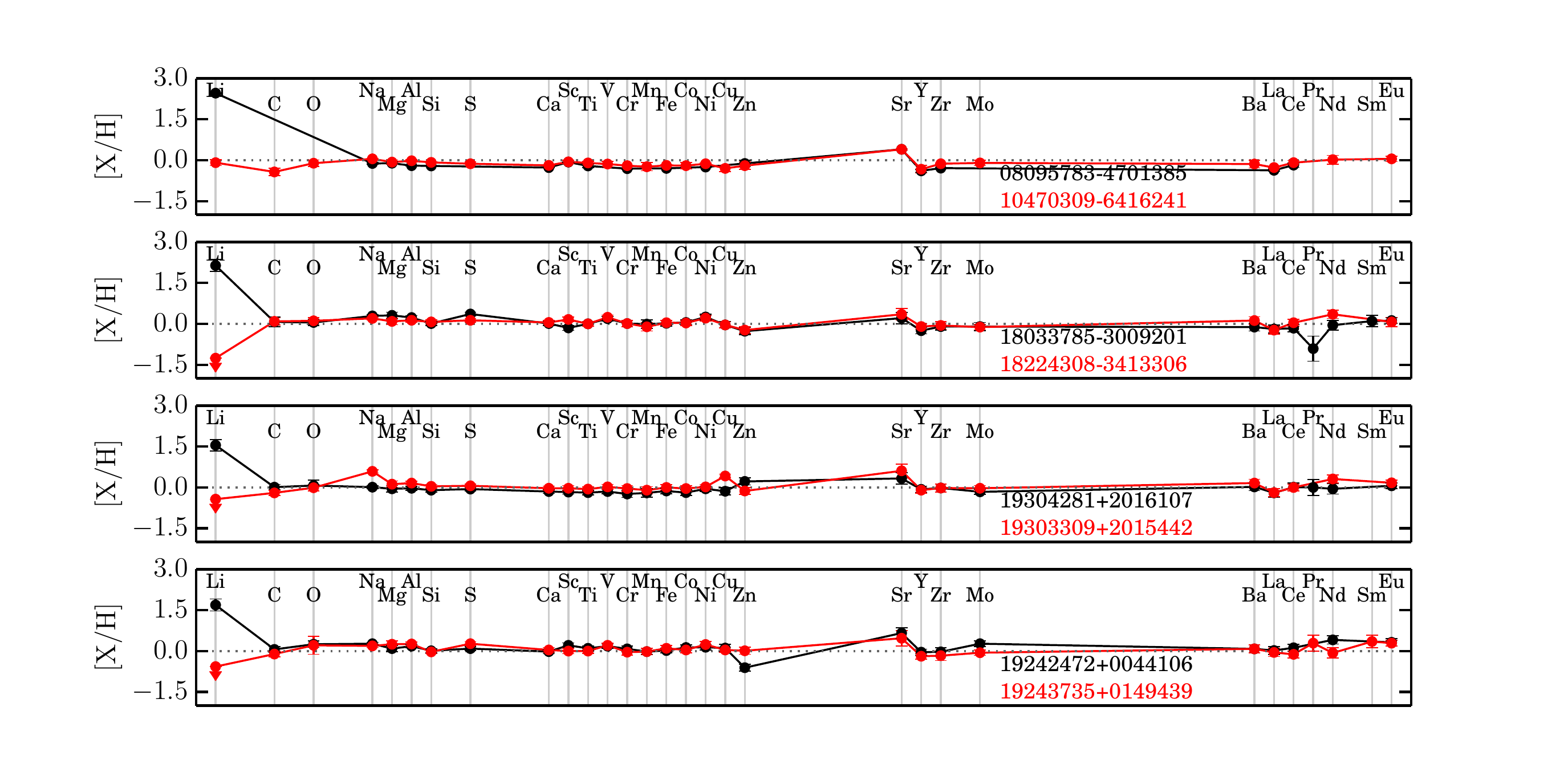}
\caption{Detailed chemical abundances for the 4 Li-rich giant stars identified in the UVES sample of the \ges\ iDR4. The detailed chemical abundances of a comparison Li-normal/poor star of similar stellar parameters is shown in red for each Li-rich giant.}
\label{fig:uves-abundances}
\end{figure*}

The \ges\ reports individual chemical abundances for up to 45 species in iDR4: 34 
elements at different ionisation stages.  These range from $Z = 3$--$63$ (Li 
to Eu) and include odd-Z, $\alpha$-, Fe-peak, as well as neutron-capture ($s$- 
and $r$-process) elements.  The resolution, wavelength coverage, and S/N of the GIRAFFE 
sample is inferior to UVES, therefore only a maximum of 15 species are available
from GIRAFFE spectra.  Given the S/N and spectral type of our Li-rich giant 
sample, for some stars we report abundances for only a few (or no) elements.
Tables \ref{tab:abundances-giraffe}-\ref{tab:lithium} contain the detailed
abundances for all Li-rich giants in our sample.  We find no obvious anomalous 
pattern in the detailed chemical abundances of our Li-rich stars (Figure 
\ref{fig:uves-abundances}).  This confirms findings from other studies that 
conclude Li seems to be the only element of difference 
\citep[e.g.,][]{Ruchti_2011,Martell_2013}.  For completeness purposes we have 
calculated non-LTE lithium abundances using the grid of corrections from 
\citet{Lind_2009a}.  These measurements are listed in Table \ref{tab:lithium},
but throughout this text all abundances refer to those calculated in LTE.

\begin{figure*}
\includegraphics[width=\textwidth]{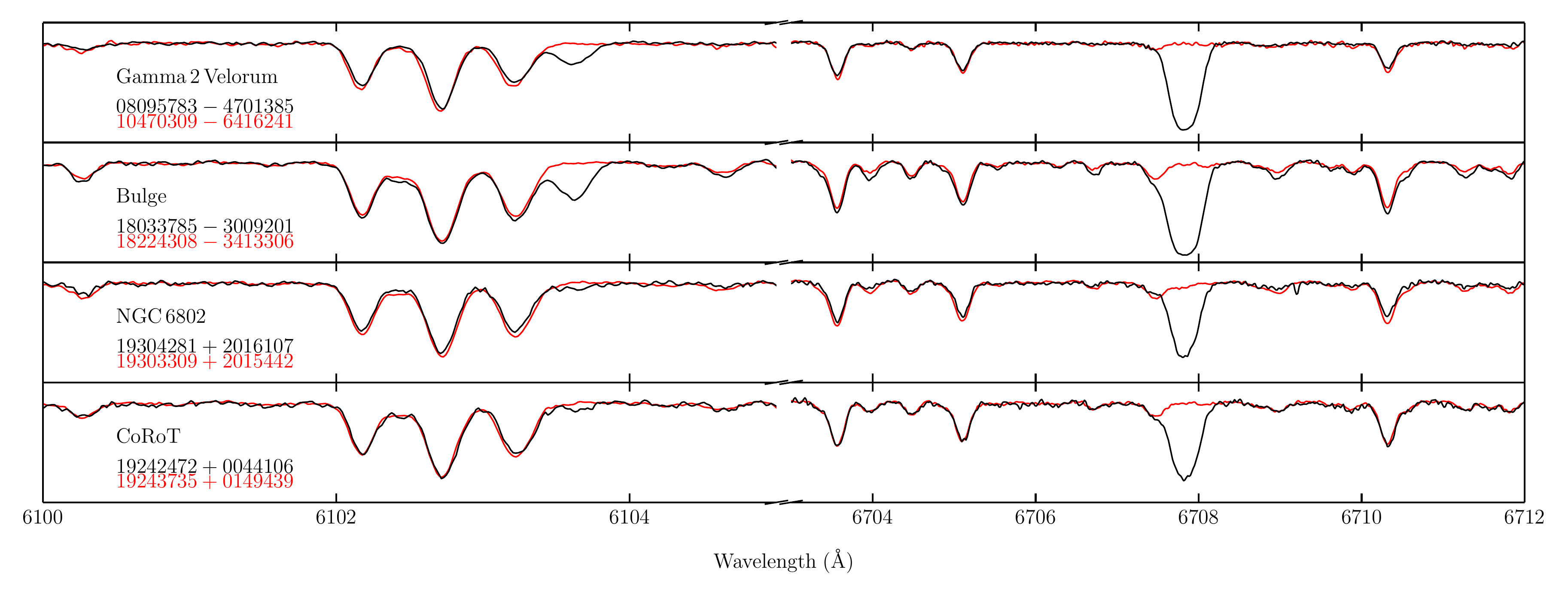}
\caption{Normalised UVES spectra of the 4 Li-rich stars identified in the \ges\ iDR4 sample (black), and
a comparison star (red) with similar stellar parameters. The resonance Li line at 6707~\AA\ and subordinate
line at 6103~\AA\ are visible, clearly showing Li enrichment.}
\label{fig:uves-spectra}
\end{figure*}

There is little doubt that these stars are indeed Li-rich.  In Figure 
\ref{fig:uves-spectra} we show the spectra surrounding the Li resonance doublet 
at 6707~\AA\ and the subordinate line at 6103~\AA\ for the Li-rich stars 
observed with UVES.  A comparison giant star of similar stellar parameters is 
shown in each panel, highlighting the difference in Li.  The 6707~\AA\ line is
strong in all four stars and saturates in the bulge star 18033785--3009201. 
The 6103~\AA\ line is also visible.  Similarly, we show the 6707~\AA\ line for 
all Li-rich stars observed with GIRAFFE in Figure \ref{fig:giraffe-spectra},
confirming their high Li abundances.  The 6103~\AA\ line is not covered by the 
GIRAFFE setups employed.

We find only one Li-rich giant star in our sample to be a fast rotator
($v\sin{i} \gtrsim 20$~km s$^{-1}$): 11000515-7623259, the star towards 
Chameleon 1.  We find no evidence of binarity in our sample: no significant 
secondary peak is seen in the cross-correlation function, and no spectral lines
are repeated.  However this does not preclude the possibility of a faint binary
companion.  Repeat radial velocity measurements over a long baseline may be 
required to infer the presence of any companion.

We searched for indications of significant mass-loss in our sample of Li-rich 
stars.  We cross-matched our sample with the Wide-Field Infrared Survey Explorer
\citep[hereafter WISE,][]{Wright_2010} and the 2 Micron All Sky Survey 
\citep[2MASS,][]{Skrutskie_2006} catalogues to search for infrared excesses that
may be attributable to ejected shells or dust-loss.  All stars had entries in 
2MASS and WISE.  We investigated all possible combinations of near- and 
mid-infrared colours and found no significant difference in the colours (or 
magnitudes) of our Li-rich stars.  Two stars exhibited mild excesses in WISE 
colours, but there are indications that the reported excess is due to source 
confusion and high background levels.  If the Li-rich stars in our sample are
experiencing significant mass-loss as dust, that signature may only be 
observable in the far infrared.  Because these stars are (relatively) faint 
(see Table \ref{tab:stellar-parameters}), they may not be visible in the far
infrared even if a substantial relative excess exists due to the presence of
a shell.

Giant stars experiencing significant mass-loss as gas often show blue-ward 
asymmetry in their H-$\alpha$ profile \citep[e.g.,][]{Meszaros_2009}.  Figure 
\ref{fig:halpha-spectra} shows spectra for all Li-rich giant stars around the
H-$\alpha$ line.  No obvious asymmetry is present for the UVES sample. 
There is some suggestion of asymmetry in some of the GIRAFFE Li-rich giants,
most notably 08102116-4740125 and 11000515-7623259.   However for most 
Li-rich stars in our sample, there is weak evidence for any recent and 
significant mass loss, either in the form of gas, dust, or shells.

\section{Discussion}
\label{sec:discussion}

The key to understanding the nature of the Li production and preservation
mechanisms in giant stars is to accurately know their evolutionary stage and the
surrounding environment. Although some of our Li-rich stars have evolved past
the RGB bump, the majority of our Li-rich giants lie just below the RGB bump.  
This is consistent with other studies of Li-rich giants of 
solar-metallicity \citep[e.g.,][and Figure \ref{fig:literature}]{Martell_2013}, whereas most metal-poor 
Li-rich giants have been found at more evolved stages: either slightly past 
the RGB bump \citep[e.g.,][]{DOrazi_2015}, towards the RGB tip, red clump, or
on the AGB \citep[e.g.,][]{Kumar_2011,Ruchti_2011}.

The fact that many of our stars lie before the RGB bump is a genuine problem,
because this is before the discontinuity in mean molecular weight can be
destroyed, irrespective of mass.  An alternative scenario is that these 
stars have simply been mis-classified as pre-bump stars 
\citep[e.g.,][]{da_Silva_2006}, and they are more likely past the luminosity 
bump or are red clump stars.

Below we discuss the observational signatures, the evolutionary stage,
environment and membership thereof for all Li-rich giant stars in our 
sample, before commenting on the plausibility of the proposed scenarios.

\subsection{Environment \& Evolution}
\subsubsection{Li-rich giants towards clusters}

Half of our Li-rich stars are in the direction of open clusters.  This is
due to an observational bias: the GIRAFFE instrument setups used for the 
\ges\ Milky Way fields do not include the Li line.  Additional setups
are used for clusters and special fields (e.g., the \corot\ fields), which include Li.
The clusters surrounding each Li-rich star
are shown in Table \ref{tab:stellar-parameters}.  Below we discuss why 
these Li-rich giant stars are unlikely to be bonafide cluster members.  
However, we stress that our conclusions are not conditional on 
(non-)membership for any of the Li-rich giant stars.  While cluster 
membership clearly has an influence on the frequency of Li-rich giant 
stars in the field and clusters (Section \ref{sec:frequency-discussion}),
these inferences are similarly complicated by the absence of quantifiable
selection functions for other Li-rich giant studies.

We find two Li-rich giants towards the young open cluster gamma2 Velorum, 
neither of which are likely members.  08102116-4740125 has a radial
velocity that is inconsistent with the cluster, and 08095783-4701385 
has a velocity near the maximum cluster value (26~km s$^{-1}$).  
More crucially, any giants towards \emph{any young cluster} like 
gamma2 Velorum (5-10~Myr) are extremely unlikely to be cluster members 
given the cluster age.  This reasoning extends to 08395152-5315159 
towards IC~2391 (53~Myr), the Li-rich giants towards NGC~2547 (35~Myr)
and Chamaeleon~1 (2~Myr), and the four Li-rich giant stars towards
IC~2602 (32~Myr).

This argument does not extend to NGC~6802, which is substantially older
(1~Gyr).  Nevertheless, the UVES Li-rich star towards NGC~6802 is also
unlikely to be a bonafide member.  \citet{Janes_Hoq_2011} classify it as a
likely non-member in their detailed cluster study, and \citet{Dias_2014} 
estimate a 66\% membership probability based on proper motions.  The 
radial velocity is mildly ($\sim{}2\sigma$) inconsistent with the
distribution of cluster velocities.  Finally, the metallicity places
19304281+2016107 a full 0.2~dex lower than the cluster mean, significantly
away from the otherwise small dispersion in metallicity seen for this
cluster.

\subsubsection{18033785-3009201, the Li-rich bulge star}
\label{sec:discussion-bulge}

The discovery of 18033785--3009201 at $(l, b) = (1^\circ, -4^\circ)$ makes it 
the most Li-rich giant star known towards the bulge \citep{McWilliam_1994,
Gonzalez_2009}.  Its radial velocity ($-70$~km~s$^{-1}$) 
is consistent with bulge membership for stars at this location \citep{Ness_2013}.

The detailed chemical abundances we derive are in excellent agreement with the
literature.  \citet{Bensby_2013} report detailed chemical abundances from 58 
microlensed dwarf and sub-giant stars in the bulge.  A comparison of their work
with respect to 18033785--3009201 is shown in Figure \ref{fig:bulge}.  Although 
we find slightly higher [Na/Fe] and [Al/Fe] ratios than \citet{Bensby_2013}, 
our abundances are consistent with other bulge studies focusing on giant stars
\citep[e.g.,][]{Fulbright_2007}.

18033785--3009201 exhibits a noteworthy deficiency in the classical $s$-process
elemental abundances: Ba, La, Ce, Pr, and Nd.  Although the 
uncertainty on Pr~II is quite large ($\sim$0.5~dex), on average we find 
18033785--3009201 to be depleted in $s$-process elements relative to iron, by 
$\sim$0.3~dex.  This signature is not seen in the classical $r$-process element 
Eu, where we find ${\rm [Eu/Fe]} = 0.05 \pm 0.10$~dex.  Low [$s$-process/Fe] 
abundance ratios are generally consistent with an ancient population (e.g., 
dwarf galaxies, however there are exceptions), and the depletion in these 
elements firmly rules out any scenarios where the increased surface Li abundance 
is associated with mass transfer from a nearby companion, which would result in
an increase of [$s$/Fe] abundance ratios.

The stellar parameters for 18033785--3009201 place it near the RGB bump.  
Given the uncertainty in $\log{g}$, we cannot rule out whether this star
is on the RGB or is actually a red clump star.  The measured [C/O] ratio of
0.03 is near-solar, and while this is only weak evidence, it suggests the star
has not completed first dredge up as a decrease in C abundances would be
expected \citep[e.g.,][]{Karakas_2014}.  A better understanding
of the evolutionary state would be useful to constrain the details of any 
internal mixing.  However we note that detecting asteroseismic oscillations 
from 18033785--3009201 is not likely in the foreseeable future, as its position
lies 2$^\circ$ from the closest planned \textit{K2}
field\footnote{http://keplerscience.arc.nasa.gov/} towards the bulge.

\subsubsection{Li-rich giants in the CoRoT field}
\label{sec:discussion-corot}

Our sample contains the first Li-rich giant stars discovered towards any \corot\ 
fields.  One star was observed with UVES, and the remaining eight using GIRAFFE. 
Most of the \corot\ Li-rich giant stars are approximately around solar 
metallicity, with a higher frequency of stars observed just below the RGB bump. 
However, at least two, perhaps three, stars are consistent with being more
evolved.

19301883--0004175 is the coolest and most metal-poor Li-rich star in our sample
($T_{\rm eff} = 4070$~K, ${\rm [Fe/H]} = -0.52$).  In contrast to observations
where most Li-rich giant stars are found below the RGB bump, 19301883--0004175
adds to the small sample of Li-rich stars at more evolved stages.  Li-rich giant 
stars past the RGB bump are preferentially more metal-poor, consistent with 
19301883--0004175.

Given the stellar parameters, 19230935+0123293 is consistent with being
a red clump star.  The uncertainties in stellar parameters for 19253819+0031094
are relatively large, therefore its exact evolutionary stage is uncertain.
Given the uncertainties in stellar parameters and the tendency of 
solar-metallicity Li-rich giants to occur more frequently around the RGB bump, 
it is perhaps likely that 19230935+0123293 is indeed located near the RGB bump, 
as indicated by the reported stellar parameters.  The ambiguity in evolutionary 
stage for these stars would be easily resolved if astereoseismic oscillations 
were detectable for these objects.  However, at this stage, it would appear 
these stars are slightly too faint for the evolutionary stage to be derived
from \corot\ light curves.

\subsection{Explaining the Li-rich giant phenomena}
\label{sec:discussion-explaining}

Here we discuss the plausibility of internal and external mechanisms proposed to
reconcile observed properties of Li-rich giants.  We note that our data are
inadequate to comment on external mechanisms involving supernovae or transient 
X-ray binaries, therefore we do not consider this hypothesis further.

The internal scenarios that we have previously outlined describe the deep 
mixing conditions required to produce an increased surface Li-abundance.  
However -- other than thermohaline mixing -- these models lack any description 
for why a given star begins to experience deep mixing, or why the frequency 
of stars undergoing deep mixing is so low.  Therefore, while the Li production 
mechanism and the conditions required for it to occur are well-understood, there
still exists a \textit{missing link} in exactly what causes the extra 
mixing.

\subsubsection{Are Li-rich K-type giants likely due to planet ingestion?}
\label{sec:discussion-planet-ingestion}

The increasing number of stars known to host close-in giant planets
(``hot Jupiters'') provides a potential solution to the Li-rich giant problem.
In this framework two factors actually contribute towards the increase in 
surface Li abundance: (1) the injection of a large planet provides a reservoir
of primordial (unburnt) levels of lithium, and (2) deep mixing that is 
induced as the planet is dissipated throughout the convective envelope, bringing
freshly produced Li to the surface.

\citet{Siess_Livio_1999a,Siess_Livio_1999b} first explored this scenario
theoretically and showed that while the results are sensitive to the accretion
rate and structure of the star, the accretion of a planet or brown dwarf star can 
produce the requisite surface Li abundance and explain their frequency.  
However, this mechanism was invoked to reconcile the existence of Li-rich giants
across the RGB and the AGB, which is not commensurate with the properties 
of close-in hot Jupiters or their occurrence rates.

Exoplanet occurrence rates are correlated with the host star.  For example, 
close-in giant planets form preferentially around metal-rich
stars \citep[e.g.,][]{Santos_2004,Fischer_2005}.  Indeed, the frequency of 
metal-rich giant planets is well-represented as a log-linear function of the 
host star metallicity \citep[e.g.,][]{Fischer_2005}.  For FGK stars with 
near-solar metallicity, the fraction of stars hosting close-in giant planets
is approximately 8\%, and decreases to 0.6\% for stars of 
${\rm [Fe/H]} = -0.5$ \citep{Schlaufman_2014}.

The occurrence rate of close-in giant planets also appears to be a function of 
the evolutionary state of the host star.  It is well-established that sub-giant
stars have systematically higher giant planet occurrence rates when all orbital
periods are considered.  However, sub-giant stars are also found to have fewer
close-in hot Jupiters than main-sequence stars of the same metallicity 
\citep{Bowler_2010,Johnson_2010}.  There has been considerable debate to explain
the differing occurrence rates of close-in hot Jupiters, including suggestions 
that stellar mass differences between the two populations is sufficient to 
explain the discrepancy \citep{Burkert_Ida_2007,Pasquini_2007,
Kennedy_Kenyon_2008a,Kennedy_Kenyon_2008b}.  If the sub-giant stellar masses 
were considerably larger than those of main-sequence stars at the same 
metallicity, then one could imagine changes in the proto-planetary disk or
dissipation timescales (due to increased radiative pressure) that could hamper 
the formation of close-in giant planets and reconcile the observations 
\citep{Kennedy_Kenyon_2009}.  The alternative scenario is that close-in giant
planets become tidally destroyed as stars leave the main-sequence and the
convective envelope increases.  It would be difficult to unambiguously resolve 
these two possibilities (difference in stellar masses or tidal destruction of
hot Jupiters) using models of stellar evolution and planet formation, given the
number of unknown variables.

\citet{Schlaufman_Winn_2013} employed a novel approach to untangle this mystery
using precise Galactic space motions.  Their sample comprised main-sequence
and sub-giant F- and G-type stars in the thin disk.  Thin disk stars form
with a very cold velocity distribution because they grow from dense, turbulent
gas in a highly dissipative process.  Over time the velocity 
distribution for a thin disk stellar population increases due to interactions
between stars, molecular clouds and spiral waves.  Because massive stars spend
very little time on the main-sequence, there is only a short period for 
interactions to kinematically heat a population of massive stars.  In contrast,
solar-mass stars spend a long time on the main-sequence, allowing for plenty of
interactions to kinematically heat the population.  For these reasons one would 
expect the space velocity dispersion of thin disk stars to decrease with 
increasing stellar mass.  This logic extends to evolved stars, since they only
spend a small fraction as a sub-giant or giant relative to their main-sequence 
lifetime.

Using precise parallaxes and proper motions from \textit{Hipparcos} 
\citep{van_Leeuwen_2007}, \citet{Schlaufman_Winn_2013} find that the 
distribution of Galactic space motions of planet-hosting sub-giant stars are on
average equal to those of planet-hosting main-sequence stars.  For this reason,
the distribution of planet-hosting sub-giant and main-sequence stars can only
differ in age (or radius, as expected from the increasing stellar envelope), but
not mass.  Moreover the orbital eccentricities of Jupiters around sub-giants
are systematically lower than those of main-sequence stars 
\citep[e.g.,][]{Jones_2014}, indicating that some
level of angular momentum transfer and orbital circularisation has occurred.
Because the main-sequence and sub-giant planet-host stars are likely to only 
differ in age, they provide insight on what happens to close-in giant planets 
when a star's convective envelope deepens at the base of the giant branch.
Therefore the lack of close-in giant planets orbiting sub-giant stars 
provides clear evidence for their destruction \citep[e.g.,][]{Rasio_1996,
Villaver_Livio_2009,Lloyd_2011,Schlaufman_2014}.

Given this empirical evidence for tidal destruction of close-in hot Jupiters as 
a star begins its ascent on the giant branch, it is intriguing to consider what
impact the planet accretion would have on the host star. 
\citet{Siess_Livio_1999b} show that while the extent of observable signatures 
are sensitive to the mass of the planet and the accretion rate, the engulfment 
of a close-in giant planet can significantly increase the photospheric Li 
abundance.  Recall that two factors contribute to this signature.  Firstly,
the accreted mass of the giant planet -- where no Li burning has occurred -- 
can produce a net increase in photospheric Li.  The second effect allows for
Li \textit{production} within the star: the spiralling infall of a giant planet
and the associated angular momentum transfer is sufficient to induce deep 
mixing, bringing freshly produced $^7$Li to the surface before it is destroyed.

If the additional Li reservoir were the only effect contributing to the net
increase in photospheric Li, then an order-of-magnitude estimate of the 
requisite planetary mass suggests a brown dwarf is required. However a brown
dwarf will have a fully convective envelope, and will therefore have depleted 
some of its primordial Li abundance.  Moreover, the lack of brown dwarfs found
within 3-5~AU around solar-mass stars \citep[$<$1\%; the \emph{`brown dwarf 
desert'}, see][]{Grether_2006} indicates that brown dwarfs are not frequent 
enough to later produce the higher frequency of Li-rich giant stars.  For 
these reasons Li-rich giant stars are unlikely to be primarily produced from
the ingestion of a brown dwarf, implying that the deep mixing induced by 
angular momentum transfer is crucial to produce high photospheric Li 
abundances.  Moreover, without any additional mixing (and just a reservoir of
unburnt Li) we would expect a similar increase in Be, which has not been 
detected in Li-rich giant stars to date \citep{De_Medeiros_1997,Castilho_1999,
Melo_2005,Monaco_2014}.

Indeed, if we simply take the models of \citet{Siess_Livio_1999a,
Siess_Livio_1999b} at face value and assume that \emph{some} conditions of accretion 
rate can produce a net increase in photospheric Li (either through a fresh
reservoir of Li and induced deep mixing), then the observed occurrence rates
of close-in giant planets \textit{predicts a population of Li-rich
giant stars before the RGB bump}.  The occurrence rates of close-in giant
planets at solar metallicity \citep[$\approx$8\%, or more conservatively 
$\approx$1\%, e.g., ][]{Santerne_2015} is commensurate with the idea that some
accretion conditions could produce a population of Li-rich giant stars with a 
frequency of $\approx$1\%.

If this scenario were true, the correlation between the occurrence rate of 
close-in giant planets and the host stellar metallicity suggests that we should
expect to see more Li-rich giant stars before the RGB bump with higher 
metallicities.  Although the lack of reproducible selection functions for 
studies of Li-rich giant stars prevents us from commenting on the fraction of 
Li-rich giants at a given metallicity, the observations are consistent with our
expectations.  Indeed, like \citet{Martell_2013}, we find that most of our 
Li-rich giant stars have near-solar metallicities.  However this observation may
be complicated by the \ges\ selection function, as the metallicity
distribution function of \ges\ stars peaks near solar metallicity for the UVES
sample in iDR4.

Contrary to the original motivation in \citet{Siess_Livio_1999a}, the planet
engulfment model is actually less likely to produce Li-rich stars all across the 
RGB and AGB, because close-in giants are likely to be destroyed as soon as the
convective envelope increases.  Although planets are found more frequently 
around sub-giant stars, those planets are preferentially found on long orbital 
periods.  Moreover, the timescale of Li-depletion suggests that our proposed 
scenario is unlikely to account for highly evolved stars with increased Li.  As
the planet is destroyed the subsequent Li enhancement will be depleted over the
next $\sim$0.2--1~Myr.  Because low-mass stars spend such a short time from the 
main-sequence to the sub-giant phase, we should expect any Li enhancement to be
depleted by the time they have ascended even moderately up the giant branch.

Alternatively, if a giant planet is formed sufficiently far from the host star
it may be unaffected by the initial expansion of the convective envelope.  In 
this scenario it may be accreted at a subsequent time, ultimately being 
destroyed when the star is more evolved.  However the circularisation and long 
orbital periods of giant planets around sub-giant stars suggests that the 
long-timescale engulfment scenario is somewhat improbable \citep{Jones_2014,Schlaufman_2014}.
On the other hand, one could imagine a somewhat unusual scenario where the planet is not fully 
dissolved, and orbits within the stellar photosphere without any large transfer
of angular momentum.  In principle, this kind of scenario may explain Li-rich 
giant stars at more evolved stages.

Our assertion linking the majority of Li-rich stars as a consequence of tidal
destruction of close-in giant planets is unlikely to fully explain the
existence of very metal-poor Li-rich giants.  The occurrence rates of 
close-in giant planets for stars with low metallicity (${\rm [Fe/H]} = -0.5$) 
is a mere $\sim$1\%, and decreases with total metallicity.  Therefore, a
very metal-poor star (e.g., ${\rm [Fe/H]} < -2$) is quite unlikely to host
any planet (including a close-in giant planet), and therefore planet accretion
is an improbable explanation for the increased surface Li.  However, of the 
Li-rich stars that are also metal-poor, these are almost ubiquitously found to 
also be highly evolved (e.g., AGB, RGB tip, red clump), which are thus
explainable through a host of internal mechanisms.

Dynamical interactions would suggest that our proposed link between close-in
giant planets and Li-rich giants implies a lower fraction of Li-rich giant stars
should be found in dense stellar environments.  Three body interactions in a 
dense cluster can sufficiently perturb a close-in hot Jupiter before a star 
leaves the main-sequence \citep{Sigurdsson_1992,Hurley_Shara_2002}.  While the 
evidence is weak, this appears to be consistent with the observations of Li-rich
giants (see Section \ref{sec:frequency-discussion}).

\subsubsection{Has the evolutionary stage been mis-estimated?}
\label{sec:discussion-evolutionary-stage}

An alternative scenario is that spectroscopic studies of Li-rich giants are
systematically biased in their determination of surface gravities.  Indeed, if
the majority of Li-rich giant stars are actually red clump stars that have been
misclassified as stars below the bump, there may be little or no requirement for 
an external mechanism to induce additional mixing.

In their low-resolution study of $\sim 2,000$ low-mass giant stars, 
\citet{Kumar_2011} identified fifteen new Li-rich stars and noted a
concentration of them at the red clump, or on the RGB.  Either evolutionary
state was plausible, as it is difficult to unambiguously determine the
precise evolutionary state directly from spectroscopy.  Because the lifetime for 
clump stars is much longer than those at the bump, it is reasonable to expect
that many field stars identified to be near the luminosity bump are indeed clump
stars.  Moreover, stellar evolution models suggest that Li can be synthesised 
during the He-core flash \citep{Eggleton_2008,Kumar_2011}, suggesting that most
Li-rich giants may actually be red clump stars, and have been mis-identified as
being near the luminosity bump.

\citet{Silva_Aguirre_2014} used asteroseismic data from the \kepler\ space 
telescope and came to this conclusion for their metal-poor 
(${\rm [Fe/H]} = -0.29$) Li-rich star.  Although stellar parameters derived from
spectroscopy alone were unable to confidently place their star on the RGB or at 
the clump, the internal oscillations for a star with or without a He-burning 
core show small differences \citep{Bedding_2011,Mosser_2011}.  However
\citet{Jofre_2015} showed that solar-like oscillations in KIC 9821622 (another 
Li-rich giant star) demonstrated that it does not have a He-burning core, 
and firmly places the evolutionary stage of KIC 9821622 below the luminosity
bump on the giant branch.

Our sample constitutes the largest number of Li-rich giant stars identified in a field 
observed by a space telescope capable of detecting asteroseimic oscillations.  
Although our stellar parameters are more consistent with the majority of these 
stars being on the RGB at or below the luminosity bump, they are each 
\textit{individually} consistent with being red clump stars: the red clump
position (in $T_{\rm eff}$ and $\log{g}$) is 1-$\sigma$ to 2-$\sigma$ of the
quoted uncertainty for each \textit{individual} star.  However as a coherent
sample, the \textit{population} significance depends on how correlated
these measurements are.  For these reasons, employing asteroseismic data from 
\corot\ may reveal whether these stars are indeed red clump stars, or associated
with the bump in the luminosity.  If indeed it is the former, an external planet
ingestion scenario becomes unlikely, which would provide strong direction on 
where to focus modelling efforts.  We encourage follow-up work to distinguish 
these possibilities.

\subsection{Frequency of Li-rich K giants}
\label{sec:frequency-discussion}

The selection function and observing strategy employed for the \ges\ preclude us
from robustly commenting on the frequency of Li-rich giants for the Milky Way field 
population.  All UVES spectra include the 6707~\AA\ Li line, but the standard 
GIRAFFE settings used for Milky Way \textit{Survey} fields (HR10 and HR21) do 
not span this region.  \corot\ observations within the \ges\ are a unique subset
of high scientific interest, which is why the HR15N setup (covering Li) was
employed for these stars.  Therefore we can only comment on the frequency of 
Li-enhanced (A(Li) $\gtrsim 2$) K-giant stars identified in the 
\corot\ field, or the fraction observed in the larger UVES sample.

At first glance the discovery of 9 Li-rich giant stars in the \corot\ field may
appear as a statistically high number, suggesting there may be something special
about the location of the \corot\ field, or the distribution of stellar masses 
within it.  The \ges\ iDR4 contains 1,175 giant stars that match our selection 
criteria ($\log{g} < 3$ and $T_{\rm eff} < 5200$~K) where the abundance of Li is 
reported.  We identify 9 Li-rich giants, resulting in an observed frequency of
slow rotating Li-rich K giants of $\sim$1\%, consistent with previous studies
\citep[e.g.,][]{Drake_2002}.

The frequency in the total UVES sample from the \ges\ iDR4 is even smaller.  The
sample contains 992 giants that match our selection criteria, of which 845 have
Li abundance measurements or upper limits.  Four of these are Li-rich, implying a 
frequency of just 0.4\%.  These are small-number statistics that may be strongly
impacted by the \textit{Survey} selection function.  For example, the UVES 
sample contains a considerable fraction (27\%) of cluster stars.  Only about 
50\% of the sample are Milky Way fields, with the remainder comprised of bulge 
fields, benchmark stars, and radial velocity standards.  The UVES cluster sample
(open and globular) contains 256 stars, of which two are Li-rich.

It is of interest to speculate whether the occurrence rate of slow-rotating 
Li-rich K giants differs between clusters and the field.  While it is difficult
for us to make robust inferences on the field frequency based on the literature or
the iDR4 \textit{Survey} data set, it is important to note that the vast 
majority ($\gtrsim90$\%) of Li-rich giant stars \textit{have} been discovered in
the field.  After accounting for the fact that star clusters have been 
extensively observed with multi-object spectroscopic instruments for over a 
decade, it seems curious that less than ten Li-rich giant stars have been detected 
in globular clusters to date.  However, we stress that standard instrumental
setups do not always include the Li line, so this line of argument is further
complicated by observational (or scientific) biases.

\section{Conclusions}
\label{sec:conclusions}

We have presented one of the largest samples of Li-rich K-giant stars.  
Our sample of Li-rich giant stars includes the most Li-rich giant known towards
the bulge, and the first sample of Li-rich giants towards the \corot\ fields.  Most 
stars have stellar parameters and abundances that are consistent with being just
below the luminosity bump on the red giant branch.  Given that about half of our 
sample is towards the \corot\ fields, accurately knowing the evolutionary stage of 
this sample could confirm their position below the luminosity bump.

The ensemble properties of Li-rich giant stars in the literature suggest two 
sub-classes, which may point towards their formation mechanism(s). 
The first is comprised of near-solar (${\rm [Fe/H]} \gtrsim -0.5$) metallicity 
stars, which are preferentially found slightly before or near the luminosity 
bump.  The second class of Li-rich giants are found in later evolutionary 
stages and are usually more metal-poor. 

We argue that Li-rich giant stars before or near the luminosity bump are a
consequence of planet/brown dwarf engulfment when the stellar photosphere 
expands at the sub-giant stage. Our assertion is supported by recent evidence on the occurrence rates of 
close-in giant planets, which demonstrate that hot Jupiters are accreted onto 
the host star as they begin to ascend the giant branch.  If we take planet accretion models at face
value and trust that \emph{some} conditions of accretion rate can produce a net 
positive abundance of Li by amassing unburnt Li \emph{and}
inducing deep mixing by angular momentum transfer, then these two lines of 
evidence actually \textit{predict the existence of Li-rich giant stars}.

This scenario would predict an increasing frequency of Li-rich giant stars with 
increasing metallicity, and the Li-depletion timescales would suggest that 
these stars should be preferentially found below the RGB bump.  Moreover, it 
would imply a lower fraction of Li-rich giant stars in dense stellar
environments (e.g., clusters) due to three body interactions.  The majority of
Li-rich giant stars are consistent with these predictions.  The remainder are 
mostly Li-rich giant stars at late evolutionary stages, a fact that is reconcilable with internal mixing prescriptions, late-time engulfment, or mass-transfer
from a binary companion.

\section*{Acknowledgments}
We thank the anonymous referee for a constructive report which improved this
paper.  We thank Ross Church, Ghina Halabi, Jarrod Hurley, Benoit Mosser, 
Kevin Schlaufman, as well as the \emph{Gaia/Stars} and \emph{Stellar Interiors}
research groups at the Institute of Astronomy, University of Cambridge for 
helpful discussions on this work.  Based on data products from observations made
with ESO Telescopes at the La Silla Paranal Observatory under programme ID 
188.B-3002.  These data products have been processed by the Cambridge Astronomy
Survey Unit (CASU) at the Institute of Astronomy, University of Cambridge, and
by the FLAMES/UVES reduction team at INAF/Osservatorio Astrofisico di Arcetri.  
These data have been obtained from the Gaia-ESO Survey Data Archive, prepared 
and hosted by the Wide Field Astronomy Unit, Institute for Astronomy, University
of Edinburgh, which is funded by the UK Science and Technology Facilities 
Council.  This work was partly supported by the European Union FP7 programme 
through ERC grant number 320360 and by the Leverhulme Trust through grant 
RPG-2012-541.  We acknowledge the support from INAF and the Ministero dell’ 
Istruzione, dell’ Università’ e della Ricerca (MIUR) in the form of the grant 
“Premiale VLT 2012” and “The Chemical and Dynamical Evolution of the Milky Way
and Local Group Galaxies” (prot. 2010LY5N2T).  The results presented here 
benefit from discussions held during the Gaia-ESO workshops and conferences 
supported by the ESF (European Science Foundation) through the GREAT Research 
Network Programme.  G.~R. acknowledges support from the project grant ``The 
New Milky Way" from the Knut and Alice Wallenberg Foundation. G.~M.~K is 
supported by the Royal Society as a Royal Society University Research Fellow. 
L.~S. acknowledges support provided by the Chilean Ministry of Economy, 
Development, and Tourism’s Millennium Science Initiative through grant 
IC120009, awarded to The Millennium Institute of Astrophysics, MAS. Funding 
for the Stellar Astrophysics Centre is provided by The Danish National 
Research Foundation. The research is supported by the ASTERISK project 
(ASTERoseismic Investigations with SONG and Kepler) funded by the European 
Research Council (Grant agreement no.: 267864). V.~S.~A. acknowledges support
from VILLUM FONDEN (research grant 10118). S.~L.~M acknowledges support from the Australian
Research Council through DECRA fellowship DE140100598. T.~M. acknowledges financial 
support from Belspo for contract PRODEX \emph{Gaia}-DPAC. J.~M. acknowledges
the support from the European Research Council Consolidator Grant funding
scheme (\emph{project STARKEY}, G.A. n. 615604). This research has 
made use of the ExoDat Database, operated at LAM-OAMP, Marseille, France, on
behalf of the CoRoT/Exoplanet program. This research made use of Astropy, a 
community-developed core Python package for Astronomy \citep{astropy}.  
This research has made use of NASA's Astrophysics Data System Bibliographic
Services.

\clearpage

\clearpage

\clearpage

\clearpage

\clearpage

\setcounter{figure}{3} 
\begin{figure*}
\centering
   \begin{subfigure}{0.45\linewidth} \centering
        \includegraphics[scale=0.6]{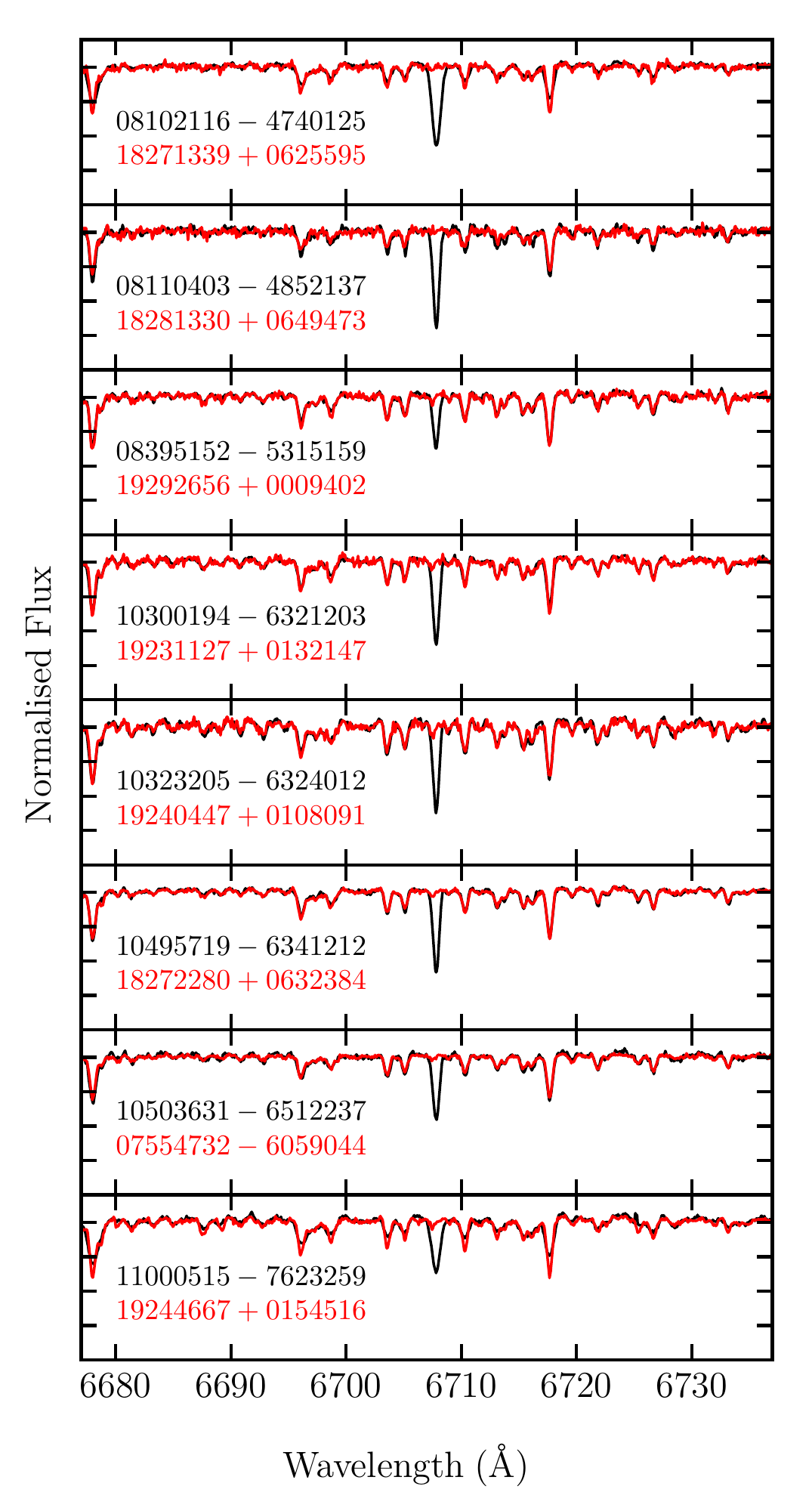}
   \end{subfigure}
   \begin{subfigure}{0.45\linewidth} \centering
        \includegraphics[scale=0.6]{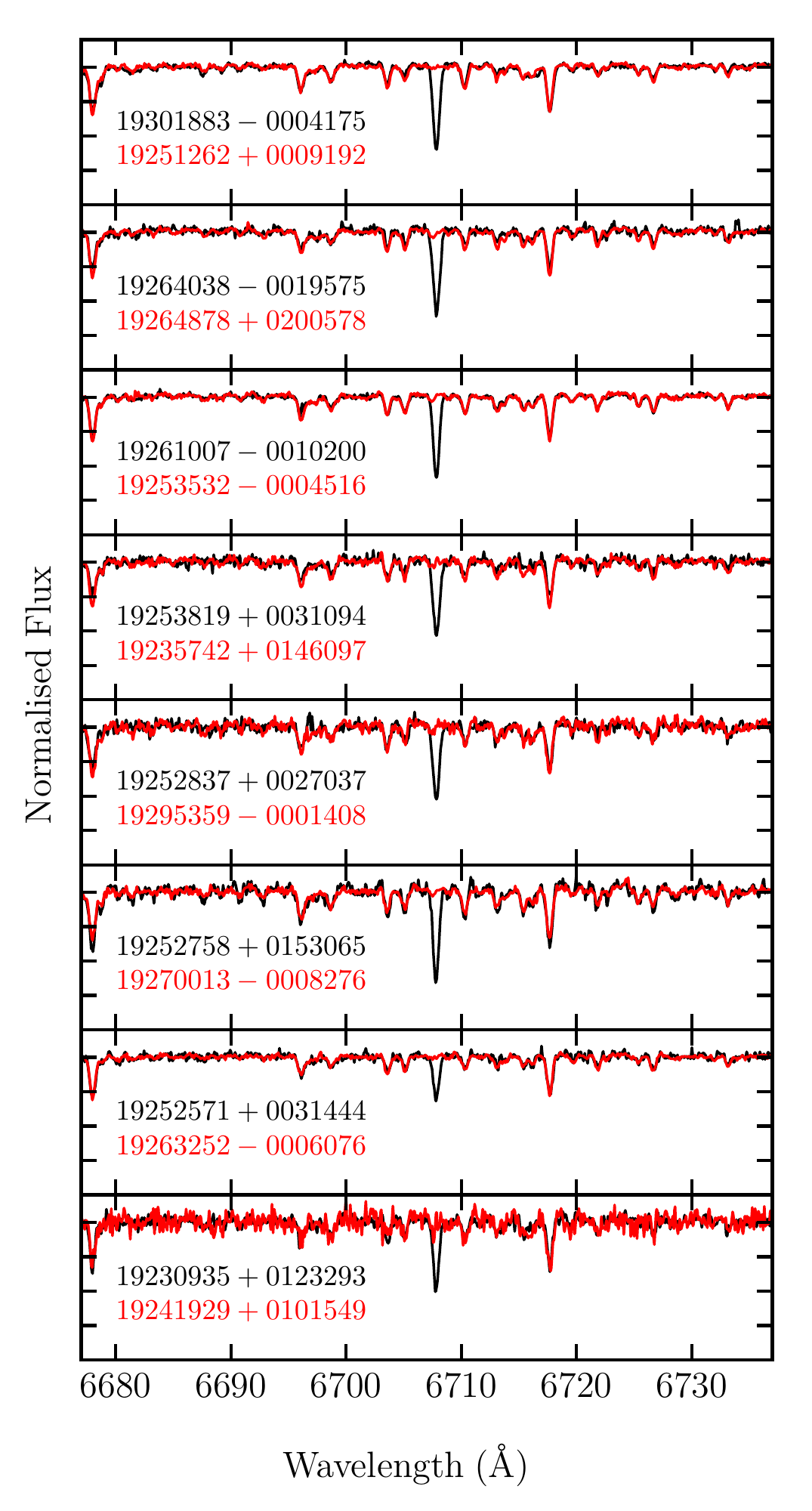}
   \end{subfigure}
\caption{Normalised spectra of the Li-rich giant stars observed using GIRAFFE. The resonance Li line at 6707~\AA\ is shown. Spectra for a Li-normal comparison star with similar stellar parameters is shown for each Li-rich giant (red).} \label{fig:giraffe-spectra}
\end{figure*}

\clearpage

\begin{figure*}
\includegraphics[width=\textwidth]{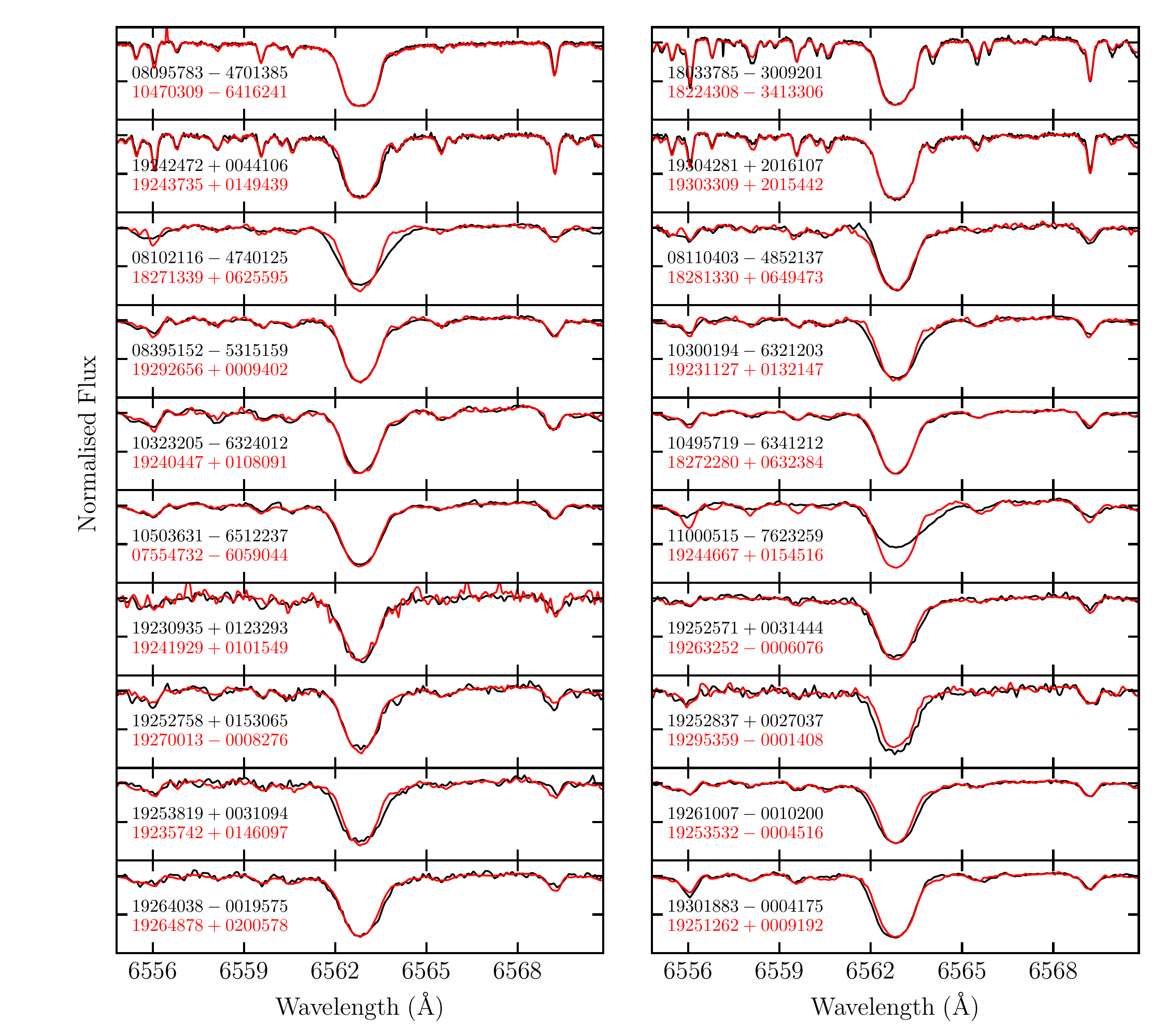}
\caption{A portion of normalised rest-frame spectra for all Li-rich giant stars (black), centered on the H-$\alpha$ line.  The UVES stars are shown in the top four panels.  If our Li-rich giant stars are experiencing mass-loss through gas, it may be apparent in asymmetries or shifts of H-$\alpha$. A comparison Li-normal/poor star is shown in red for each Li-rich giant.}
\label{fig:halpha-spectra}
\end{figure*}

\clearpage

\begin{figure}
\includegraphics[width=8.3cm]{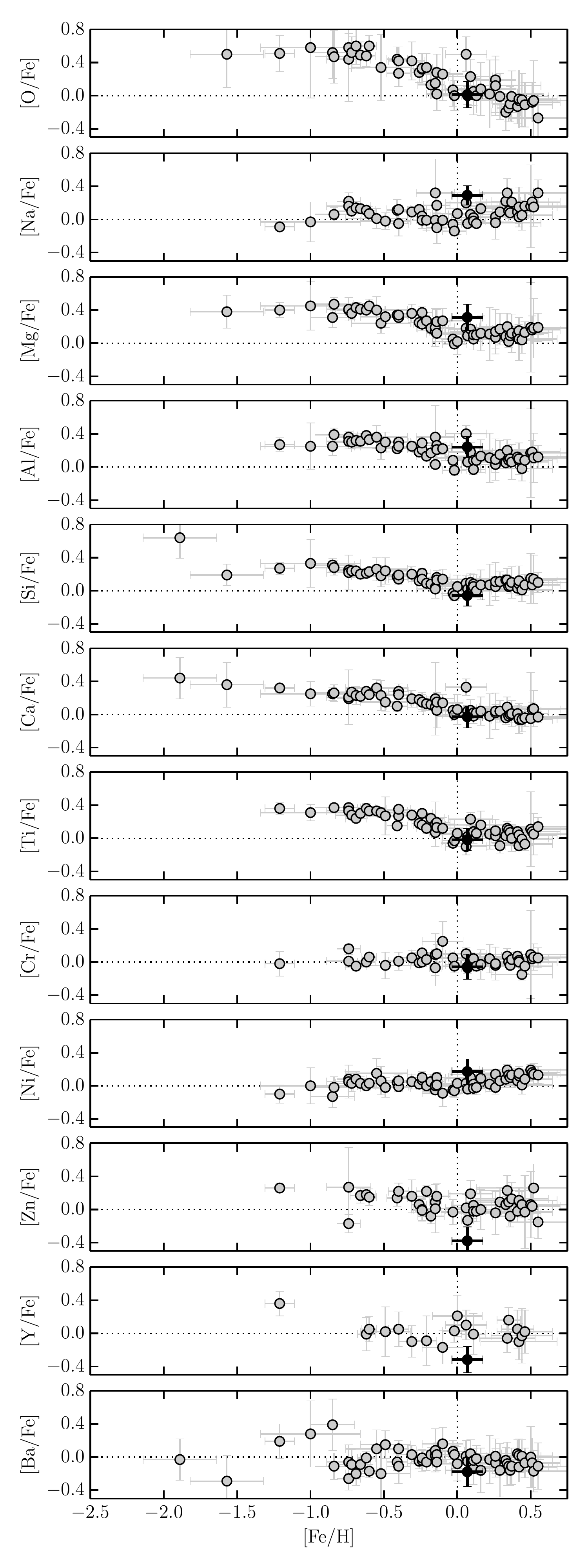}
\caption{Detailed chemical abundances of the Li-rich bulge star, 18033785$-$3009201, compared
to the microlensed bulge dwarf and sub-giant sample of \citet{Bensby_2013}.  The uncertainty in a given [X/Fe] abundance ratio for 18033785$-$3009201 is taken as the quadrature sum of [X/H] and [X/Fe].}
\label{fig:bulge}
\end{figure}

\clearpage

\begin{table*}
\begin{tabular}{clccccccl}
\hline
& Object & $T_{\rm eff}$ & $\log{g}$ & [Fe/H] & A(Li) & $v\sin{i}$ & Year & Reference \\
\hline
1 & HD172365 & 5500 & 2.1 & -0.6 & 2.49 & 70 & 1982 & \citet{Luck_1982} \\
2 & HD 174104 & 5750 & 0.9 & -0.3 & 3.46 & 50 & 1982 & \citet{Luck_1982} \\
3 & HD 112127 & 4750 & 2.6 & 0.3 & 3.2 & \nd & 1982 & \citet{Wallerstein_Sneden_1982} \\
4 & 9 Bootis (BS 5247) & 4000 & 2.0 & 0.1 & 2.5 & \nd & 1984 & \citet{Hanni_1984} \\
5 & NGC7789-443 & 5600 & 3.1 & \nd & 2.4 & 44 & 1986 & \citet{Pilachowski_1986} \\
6 & NGC7789-1238 & 5800 & 3.1 & \nd & 2.4 & $<$10 & 1986 & \citet{Pilachowski_1986} \\
7 & NGC7789-308 & 6350 & 3.3 & \nd & 3.0 & 80 & 1986 & \citet{Pilachowski_1986} \\
8 & NGC7789-268 & 6450 & 3.4 & \nd & 3.3 & 30 & 1986 & \citet{Pilachowski_1986} \\
9 & HD 183492 & 4700 & 2.4 & 0.08 & 2.0 & \nd & 1989 & \citet{Brown_1989} \\
10 & HD 126868 & 5440 & 3.2 & -0.25 & 2.3 & \nd & 1989 & \citet{Brown_1989} \\
11 & HD 112127 & 4340 & 2.1 & 0.31 & 2.7 & \nd & 1989 & \citet{Brown_1989} \\
12 & HD 108471 & 4980 & 2.8 & -0.02 & 2.0 & \nd & 1989 & \citet{Brown_1989} \\
13 & HD 148293 & 4640 & 2.5 & 0.23 & 2.0 & \nd & 1989 & \citet{Brown_1989} \\
14 & HD 9746 & 4420 & 2.3 & -0.13 & 2.7 & \nd & 1989 & \citet{Brown_1989} \\
15 & HD 39853 & 3900 & 1.16 & -0.5 & 2.8 & \nd & 1989 & \citet{Gratton_1989} \\
16 & Be21-T33 & 4600 & 2.0 & -0.58 & 3.0 & \nd & 1999 & \citet{Hill_1999} \\
17 & HD 219025 & 4500 & 2.3 & -0.1 & 3.0 & 23 & 1999 & \citet{Jasniewicz_1999} \\
18 & HDE 233517 & 4475 & 2.25 & -0.37 & 3.85 & 17.6 & 2000 & \citet{Balachandran_2000} \\
19 & HD 9746$^a$ & 4400 & 2.3 & \nd & 3.44 & 9 & 2000 & \citet{Balachandran_2000} \\
20 & HD 172481 & 7250 & 1.5 & -0.55 & 3.57 & 14 & 2001 & \citet{Reyniers_Van_Winckel_2001} \\
\hline
\end{tabular}
\caption{Reported parameters of all known Li-rich giant stars. Only a portion of the table is shown here. The full compilation is available in the online journal. \textbf{Before acceptance, full data available at:} https://docs.google.com/spreadsheets/d/1fiHZXyQrGXJivZJh8GCdwc2Zf7wepsTUHBR8BiDfwaI/edit?usp=sharing
}
\end{table*}

\clearpage

\begin{table*}
\begin{tabular}{llccccrccr}
\hline
Star & Field & $\alpha$ & $\delta$ & $J$ & $K$ & $V_{rad}$ & $T_{\rm eff}$ & $\log{g}$ & [Fe/H] \\
 & & (J2000) & (J2000) & & & & (km~s$^{-1}$) & (K) & \\
\hline
08095783$-$4701385 & $\gamma$2 Velorum  & 08:09:57.83 & $-$47:01:38.5 & 10.5 &  9.8 &  $25.6 \pm 0.2$ & 4964 & 2.43 & $-0.15$  \\
18033785$-$3009201 & Bulge              & 18:03:37.85 & $-$30:09:20.1 & 11.4 & 10.6 & $-70.0 \pm 0.6$ & 4467 & 2.34 &  $0.07$  \\
19242472$+$0044106 & \corot\            & 19:24:24.73 & $+$00:44:10.5 & 11.3 & 10.5 &  $77.7 \pm 0.1$ & 4740 & 2.70 &  $0.08$  \\
19304281$+$2016107 & NGC~6802           & 19:30:42.81 & $+$20:16:10.7 & 11.7 & 10.7 &  $17.4 \pm 0.6$ & 4766 & 2.63 & $-0.10$  \\
\hline
08102116$-$4740125 & $\gamma$2 Velorum  & 08:10:21.16 & $-$47:40:12.5 & 11.5 & 10.6 &  $71.0 \pm 0.2$ & 4591 & 2.27 & $-0.12$  \\
08110403$-$4852137 & NGC~2547           & 08:11:04.03 & $-$48:52:13.7 & 12.4 & 11.6 &  $54.1 \pm 0.2$ & 4762 & 2.59 & $-0.12$  \\
08395152$-$5315159 & IC2391             & 08:39:51.52 & $-$53:15:15.9 & 12.4 & 11.5 &  $27.0 \pm 0.2$ & 4726 & 2.55 &  $0.01$  \\
10300194$-$6321203 & IC2602             & 10:30:01.94 & $-$63:21:20.3 & 11.4 & 10.6 & $-10.2 \pm 0.2$ & 4612 & 2.37 & $-0.06$  \\
10323205$-$6324012 & IC2602             & 10:32:32.05 & $-$63:24:01.2 & 11.3 & 10.6 &  $13.3 \pm 0.2$ & 4607 & 2.53 &  $0.13$  \\
10495719$-$6341212 & IC2602             & 10:49:57.19 & $-$63:41:21.2 & 11.1 & 10.3 &  $13.8 \pm 0.2$ & 4789 & 2.55 &  $0.03$  \\
10503631$-$6512237 & IC2602             & 10:50:36.31 & $-$65:12:23.7 & 11.7 & 10.8 & $-34.1 \pm 0.2$ & 4708 & 2.49 & $-0.05$  \\
11000515$-$7623259 & Chameleon 1        & 11:00:05.15 & $-$76:23:25.9 & 10.1 &  9.1 & $-15.9 \pm 0.2$ & 4505 & 2.22 &  $0.06$  \\
19230935$+$0123293 & \corot\            & 19:23:09.35 & $+$01:23:29.3 & 13.1 & 12.3 &  $11.9 \pm 0.2$ & 4845 & 2.37 & $-0.12$ \\
19252571$+$0031444 & \corot\            & 19:25:25.71 & $+$00:31:44.4 & 12.7 & 11.9 & $-38.6 \pm 0.3$ & 4825 & 2.87 & $-0.10$ \\
19252758$+$0153065 & \corot\            & 19:25:27.58 & $+$01:53:06.5 & 11.3 & 10.5 &  $28.2 \pm 0.1$ & 4617 & 2.80 & $0.21$  \\
19252837$+$0027037 & \corot\            & 19:25:28.37 & $+$00:27:03.7 & 13.4 & 12.6 &   $0.3 \pm 0.3$ & 4731 & 2.91 & $0.18$  \\
19253819$+$0031094 & \corot\            & 19:25:38.19 & $+$00:31:09.4 & 13.0 & 12.1 &  $26.6 \pm 0.3$ & 4655 & 2.51 & $-0.25$ \\
19261007$-$0010200 & \corot\            & 19:26:10.07 & $-$00:10:20.0 & 11.8 & 11.1 & $-21.6 \pm 0.2$ & 4752 & 2.84 & $0.12$  \\
19264038$-$0019575 & \corot\            & 19:26:40.38 & $-$00:19:57.5 & 13.0 & 12.2 &  $41.8 \pm 0.3$ & 4782 & 2.91 & $0.02$  \\
19301883$-$0004175 & \corot\            & 19:30:18.83 & $-$00:04:17.5 & 11.6 & 10.5 &  $57.3 \pm 0.1$ & 4070 & 1.63 & $-0.52$ \\
\hline
\end{tabular}
\caption{Positions, photometry, velocities and stellar parameters for all Li-rich stars in the sample. 
Candidates observed with UVES are at the head of the table, separated from the GIRAFFE spectra by the horizontal line.
}
\label{tab:stellar-parameters}
\end{table*}

\clearpage

\begin{table}
\begin{tabular}{llrcrr}
\hline
Element & Ion & A(X) & $\sigma$ & [X/H] & [X/Fe] \\
\hline
\multicolumn{6}{c}{08110403-4852137} \\
\hline
Ti & 1 & $4.82$ &  \nd & $-0.08$ &  $0.04$ \\
Co & 1 & $4.82$ &  \nd & $-0.10$ &  $0.02$ \\
\hline
\multicolumn{6}{c}{19230935+0123293} \\
\hline
Al & 1 & $6.48$ & 0.05 &  $0.11$ &  $0.23$ \\
Si & 1 & $7.61$ & 0.04 &  $0.10$ &  $0.22$ \\
Ca & 1 & $6.12$ & 0.10 & $-0.19$ & $-0.07$ \\
Ti & 1 & $4.94$ &  \nd &  $0.04$ &  $0.16$ \\
Co & 1 & $4.77$ &  \nd & $-0.15$ & $-0.03$ \\
Ni & 1 & $6.19$ & 0.03 & $-0.04$ &  $0.08$ \\
Ba & 2 & $1.81$ & 0.25 & $-0.36$ & $-0.24$ \\
\hline
\multicolumn{6}{c}{19252571+0031444} \\
\hline
Al & 1 & $6.29$ & 0.12 & $-0.08$ & $0.02$  \\
Si & 1 & $7.36$ & 0.02 & $-0.15$ & $-0.05$ \\
Ca & 1 & $5.97$ & 0.01 & $-0.34$ & $-0.24$ \\
Ti & 1 & $4.71$ & \nd  & $-0.19$ & $-0.09$ \\
Co & 1 & $4.75$ & \nd  & $-0.17$ & $-0.07$ \\
Ni & 1 & $5.82$ & 0.02 & $-0.41$ & $-0.31$ \\
Ba & 2 & $2.20$ & 0.08 & $0.03$  & $0.13$  \\
\hline
\multicolumn{6}{c}{19252758+0153065} \\
\hline
Mg & 1 & $7.84$ & 0.01 & $0.31$ & $0.10$  \\
Al & 1 & $6.68$ & 0.03 & $0.31$ & $0.10$  \\
Si & 1 & $7.73$ & 0.03 & $0.22$ & $0.01$  \\
Ti & 1 & $5.10$ & \nd  & $0.20$ & $-0.01$ \\
Mn & 1 & $5.48$ & 0.04 & $0.09$ & $-0.12$ \\
Fe & 1 & $7.71$ & 0.02 & $0.26$ & $0.05$  \\
Co & 1 & $4.92$ & \nd  & $0.00$ & $-0.21$ \\
\hline
\multicolumn{6}{c}{19252837+0027037} \\
\hline
Ti & 1 & $4.86$ & \nd  & $-0.04$ & $-0.22$ \\
Co & 1 & $4.93$ & \nd  & $0.01$  & $-0.17$ \\
\hline
\multicolumn{6}{c}{19253819+0031094} \\
\hline
Ti & 1 & $4.69$ & \nd  & $-0.21$ & $0.04$  \\
Co & 1 & $4.52$ & \nd  & $-0.40$ & $-0.15$ \\
Ba & 2 & $1.84$ & 0.11 & $-0.33$ & $-0.08$ \\
\hline
\multicolumn{6}{c}{19261007-0010200} \\
\hline
Ti & 1 & $4.64$ & \nd  & $-0.26$ & $-0.38$ \\
Co & 1 & $4.77$ & \nd  & $-0.15$ & $-0.27$ \\
\hline
\multicolumn{6}{c}{19264038-0019575} \\
\hline
Ti & 1 & $4.65$ & \nd  & $-0.25$ & $-0.27$ \\
Co & 1 & $4.78$ & \nd  & $-0.14$ & $-0.16$ \\
\hline
\multicolumn{6}{c}{19301883-0004175} \\
\hline
Mg & 1 & $7.48$ & 0.01 & $-0.05$ & $0.47$  \\
Al & 1 & $6.23$ & 0.01 & $-0.14$ & $0.38$  \\
Si & 1 & $7.13$ & 0.07 & $-0.38$ & $0.14$  \\
Ca & 2 & $6.12$ & 0.07 & $-0.19$ & $0.33$  \\
Ti & 1 & $4.68$ & 0.02 & $-0.22$ & $0.30$  \\
Ti & 2 & $4.81$ & 0.06 & $-0.09$ & $0.43$  \\
Cr & 1 & $5.13$ & 0.06 & $-0.51$ & $0.01$  \\
Mn & 1 & $4.77$ & 0.19 & $-0.62$ & $-0.10$ \\
Fe & 1 & $7.02$ & 0.02 & $-0.43$ & $0.09$  \\
Co & 1 & $4.47$ & 0.02 & $-0.45$ & $0.07$  \\
\hline
\end{tabular}
\caption{Chemical abundances (except Li, see Table \ref{tab:lithium}) for all Li-rich stars observed with GIRAFFE. Note that seven stars observed with GIRAFFE have no detailed chemical abundances available.}
\label{tab:abundances-giraffe}
\end{table}

\begin{table*}
\tiny
\begin{tabular}{llccrrlccrrlccrr}
\hline
\multicolumn{2}{l}{Species} & 
A(X) & $\sigma$ & [X/H] & [X/Fe] & &
A(X) & $\sigma$ & [X/H] & [X/Fe] \\
\hline

& & 
\multicolumn{4}{c}{$08095783-4701385$} & &
\multicolumn{4}{c}{$18033785-3009201$} \\

\hline
C   & 1 & \nd    & \nd  & \nd     & \nd     &&   $8.50$  & 0.18  & $0.11$    &  $0.04$    \\  
O   & 1 & \nd    & \nd  & \nd     & \nd     &&   $8.74$  & 0.12  & $0.08$    &  $0.01$    \\  
Na  & 1 & $6.12$ & 0.03 & $-0.05$ & $0.10$  &&   $6.53$  & 0.05  & $0.36$    &  $0.29$    \\  
Mg  & 1 & $7.49$ & 0.04 & $-0.04$ & $0.11$  &&   $7.91$  & 0.12  & $0.38$    &  $0.31$    \\  
Al  & 1 & $6.25$ & 0.01 & $-0.12$ & $0.03$  &&   $6.68$  & 0.07  & $0.31$    &  $0.24$    \\  
Si  & 1 & $7.30$ & 0.01 & $-0.21$ & $-0.06$ &&   $7.52$  & 0.07  & $0.01$    & $-0.06$   \\  
S   & 1 & \nd    & \nd  & \nd     & \nd     &&   $7.48$  & 0.07  & $0.34$    &  $0.27$    \\  
Ca  & 1 & $6.07$ & 0.01 & $-0.24$ & $-0.09$ &&   $6.35$  & 0.08  & $0.04$    & $-0.03$   \\  
Sc  & 1 & \nd    & \nd  & \nd     & \nd     &&   $3.00$  & 0.09  & $-0.17$   & $-0.24$   \\  
Sc  & 2 & $3.08$ & 0.02 & $-0.09$ & $0.06$  &&   $3.29$  & 0.08  & $0.12$    &  $0.05$    \\  
Ti  & 1 & $4.74$ & 0.06 & $-0.16$ & $-0.01$ &&   $4.95$  & 0.08  & $0.05$    & $-0.02$   \\  
Ti  & 2 & $4.73$ & 0.02 & $-0.17$ & $-0.02$ &&   $5.01$  & 0.09  & $0.11$    &  $0.04$    \\  
V   & 1 & \nd    & \nd  & \nd     & \nd     &&   $4.12$  & 0.08  & $0.12$    &  $0.05$    \\  
Cr  & 1 & $5.33$ & 0.03 & $-0.31$ & $-0.16$ &&   $5.65$  & 0.11  & $0.01$    & $-0.06$   \\  
Cr  & 2 & $5.35$ & 0.14 & $-0.29$ & $-0.14$ &&   $5.88$  & 0.12  & $0.24$    &  $0.17$    \\  
Mn  & 1 & \nd    & \nd  & \nd     & \nd     &&   $5.42$  & 0.16  & $0.03$    & $-0.04$   \\  
Fe  & 1 & $7.20$ & 0.02 & $-0.25$ & $-0.10$ &&   $7.52$  & 0.10  & $0.07$    &  $0.00$    \\  
Fe  & 2 & $7.17$ & 0.06 & $-0.28$ & $-0.13$ &&   $7.56$  & 0.09  & $0.11$    &  $0.04$    \\  
Co  & 1 & \nd    & \nd  & \nd     & \nd     &&   $5.04$  & 0.10  & $0.12$    &  $0.05$    \\  
Ni  & 1 & $5.97$ & 0.02 & $-0.26$ & $-0.11$ &&   $6.47$  & 0.11  & $0.24$    &  $0.17$    \\  
Cu  & 1 & \nd    & \nd  & \nd     & \nd     &&   $4.15$  & 0.12  & $-0.06$   & $-0.13$   \\  
Zn  & 1 & $4.44$ & 0.12 & $-0.16$ & $-0.01$ &&   $4.29$  & 0.13  & $-0.31$   & $-0.38$   \\  
Sr  & 1 & $3.27$ & 0.03 &  $0.35$ &  $0.50$ &&   $3.08$  & 0.21  & $0.16$    &  $0.09$    \\  
Y   & 2 & $1.82$ & 0.06 & $-0.39$ & $-0.24$ &&   $1.96$  & 0.12  & $-0.25$   & $-0.32$   \\  
Zr  & 1 & $2.29$ & 0.02 & $-0.29$ & $-0.14$ &&   $2.48$  & 0.13  & $-0.10$   & $-0.17$   \\  
Zr  & 2 & $2.37$ & 0.06 & $-0.21$ & $-0.06$ &&   $2.55$  & 0.17  & $-0.03$   & $-0.10$   \\  
Mo  & 1 & \nd    & \nd  & \nd     & \nd     &&   $1.78$  & 0.15  & $-0.14$   & $-0.21$   \\  
Ba  & 2 & \nd    & \nd  & \nd     & \nd     &&   $2.06$  & 0.14  & $-0.11$   & $-0.18$   \\  
La  & 2 & $0.73$ & 0.03 & $-0.40$ & $-0.25$ &&   $0.91$  & 0.15  & $-0.22$   & $-0.29$   \\  
Ce  & 2 & $1.40$ & 0.08 & $-0.30$ & $-0.15$ &&   $1.42$  & 0.14  & $-0.28$   & $-0.35$   \\  
Pr  & 2 & \nd    & \nd  & \nd     & \nd     &&   $-0.19$ & 0.46  & $-0.77$   & $-0.84$   \\  
Nd  & 2 & \nd    & \nd  & \nd     & \nd     &&   $1.37$  & 0.18  & $-0.08$   & $-0.15$   \\  
Eu  & 2 & \nd    & \nd  & \nd     & \nd     &&   $0.64$  & 0.10  & $0.12$    &  $0.05$    \\  
\hline
& &
\multicolumn{4}{c}{$19242472+0044106$} & &
\multicolumn{4}{c}{$19304281+2016107$} \\
\hline
C   & 1 &  $8.47$  & 0.13  & $0.08$    & $0.00$    &&  $8.20$  & 0.13  & $-0.19$   & $-0.09$   \\
N &(CN) &  $8.26$  & 0.10  & $0.48$    & $0.40$    &&   \nd    & \nd   &  \nd      & \nd       \\
O   & 1 &  $8.94$  & 0.12  & $0.28$    & $0.20$    &&  $8.76$  & 0.20  & $0.10$    & $0.20$    \\
Na  & 1 &  $6.51$  & 0.05  & $0.34$    & $0.26$    &&  $6.25$  & 0.05  & $0.08$    & $0.18$    \\
Mg  & 1 &  $7.69$  & 0.12  & $0.16$    & $0.08$    &&  $7.55$  & 0.12  & $0.02$    & $0.12$    \\
Al  & 1 &  $6.63$  & 0.07  & $0.26$    & $0.18$    &&  $6.42$  & 0.07  & $0.05$    & $0.15$    \\
Si  & 1 &  $7.52$  & 0.07  & $0.01$    & $-0.07$   &&  $7.41$  & 0.07  & $-0.10$   & $0.00$    \\
S   & 1 &  $7.21$  & 0.07  & $0.07$    & $-0.01$   &&  $7.06$  & 0.07  & $-0.08$   & $0.02$    \\
Ca  & 1 &  $6.32$  & 0.08  & $0.01$    & $-0.07$   &&  $6.19$  & 0.07  & $-0.12$   & $-0.02$   \\
Sc  & 1 &  $3.36$  & 0.09  & $0.19$    & $0.11$    &&  $2.98$  & 0.07  & $-0.19$   & $-0.09$   \\
Sc  & 2 &  $3.22$  & 0.06  & $0.05$    & $-0.03$   &&  $3.18$  & 0.07  & $0.01$    & $0.11$    \\
Ti  & 1 &  $5.05$  & 0.08  & $0.15$    & $0.07$    &&  $4.76$  & 0.08  & $-0.14$   & $-0.04$   \\
Ti  & 2 &  $5.04$  & 0.07  & $0.14$    & $0.06$    &&  $4.88$  & 0.09  & $-0.02$   & $0.08$    \\
V   & 1 &  $4.11$  & 0.08  & $0.11$    & $0.03$    &&  $3.78$  & 0.08  & $-0.22$   & $-0.12$   \\
Cr  & 1 &  $5.72$  & 0.10  & $0.08$    & $0.00$    &&  $5.40$  & 0.13  & $-0.24$   & $-0.14$   \\
Cr  & 2 &  $5.70$  & 0.14  & $0.06$    & $-0.02$   &&  $5.72$  & 0.15  & $0.08$    & $0.18$    \\
Mn  & 1 &  $5.41$  & 0.11  & $0.02$    & $-0.06$   &&  $5.23$  & 0.16  & $-0.16$   & $-0.06$   \\
Fe  & 1 &  $7.52$  & 0.09  & $0.07$    & $-0.01$   &&  $7.37$  & 0.10  & $-0.08$   & $0.02$    \\
Fe  & 2 &  $7.47$  & 0.09  & $0.02$    & $-0.06$   &&  $7.48$  & 0.08  & $0.03$    & $0.13$    \\
Co  & 1 &  $5.12$  & 0.10  & $0.20$    & $0.12$    &&  $4.80$  & 0.10  & $-0.12$   & $-0.02$   \\
Ni  & 1 &  $6.36$  & 0.10  & $0.13$    & $0.05$    &&  $6.17$  & 0.10  & $-0.06$   & $0.04$    \\
Cu  & 1 &  $4.30$  & 0.14  & $0.09$    & $0.01$    &&  $4.05$  & 0.14  & $-0.16$   & $-0.06$   \\
Zn  & 1 &  $3.95$  & 0.13  & $-0.65$   & $-0.73$   &&  $4.78$  & 0.13  & $0.18$    & $0.28$    \\
Sr  & 1 &  $3.53$  & 0.20  & $0.61$    & $0.53$    &&  $3.20$  & 0.21  & $0.28$    & $0.38$    \\
Y   & 2 &  $2.17$  & 0.11  & $-0.04$   & $-0.12$   &&  $2.14$  & 0.12  & $-0.07$   & $0.03$    \\
Zr  & 1 &  $2.55$  & 0.15  & $-0.03$   & $-0.11$   &&  $2.56$  & 0.14  & $-0.02$   & $0.08$    \\
Zr  & 2 &  $2.99$  & 0.10  & $0.41$    & $0.33$    &&  $2.90$  & 0.10  & $0.32$    & $0.42$    \\
Mo  & 1 &  $2.15$  & 0.10  & $0.23$    & $0.15$    &&  $1.72$  & 0.10  & $-0.20$   & $-0.10$   \\
Ba  & 2 &  $2.26$  & 0.14  & $0.09$    & $0.01$    &&  $2.20$  & 0.14  & $0.03$    & $0.13$    \\
La  & 2 &  $1.12$  & 0.15  & $-0.01$   & $-0.09$   &&  $0.90$  & 0.16  & $-0.23$   & $-0.13$   \\
Ce  & 2 &  $1.70$  & 0.13  & $0.00$    & $-0.08$   &&  $1.59$  & 0.15  & $-0.11$   & $-0.01$   \\
Pr  & 2 &  \nd     & \nd   & \nd       & \nd       &&  $0.72$  & 0.29  & $0.14$    & $0.24$    \\
Nd  & 2 &  $1.83$  & 0.15  & $0.38$    & $0.30$    &&  $1.37$  & 0.19  & $-0.08$   & $0.02$    \\
Eu  & 2 &  $0.84$  & 0.13  & $0.32$    & $0.24$    &&  $0.58$  & 0.10  & $0.06$    & $0.16$    \\
\hline
\end{tabular}
\caption{Detailed chemical abundances (except Li; see Table \ref{tab:lithium}) for all Li-rich stars observed with UVES.}
\label{tab:abundances-uves}
\end{table*}

\clearpage

\begin{table}
\begin{tabular}{lcc}
\hline
Star & A(Li, LTE) & A(Li, nLTE) \\
\hline
08095783$-$4701385 & 3.51 & 3.21 \\ 
18033785$-$3009201 & 3.19 & 3.11 \\
19242472$+$0044106 & 2.74 & 2.72 \\
19304281$+$2016107 & 2.60 & 2.60 \\
\hline
08102116$-$4740125 & 3.52 & 3.33 \\
08110403$-$4852137 & 3.51 & 3.25 \\
08395152$-$5315159 & 2.15 & 2.28 \\
10300194$-$6321203 & 2.96 & 2.88 \\
10323205$-$6324012 & 3.07 & 2.98 \\
10495719$-$6341212 & 3.05 & 2.94 \\
10503631$-$6512237 & 2.59 & 2.61 \\
11000515$-$7623259 & 2.59 & 2.64 \\
19230935$+$0123293 & 2.80 & 2.75 \\
19252571$+$0031444 & 2.10 & 2.22 \\
19252758$+$0153065 & 2.99 & 2.92 \\
19252837$+$0027037 & 2.86 & 2.82 \\
19253819$+$0031094 & 2.99 & 2.85 \\
19261007$-$0010200 & 2.95 & 2.88 \\
19264038$-$0019575 & 3.35 & 3.13 \\
19301883$-$0004175 & 2.52 & 2.43 \\
\hline
\end{tabular}
\caption{Non-LTE Li calculated using corrections from \citet{Lind_2009a}. Stars observed using UVES and GIRAFFE are separated by the horizontal line. For these calculations we adopted ${\xi = 1.5}$~km~s$^{-1}$ for the GIRAFFE spectra as $\xi$ measurements were unavailable.}
\label{tab:lithium}
\end{table}

\clearpage
\noindent{}$^1$Institute of Astronomy, University of Cambridge, Madingley Road, Cambridge CB3 0HA, UK\\
$^2$Lund Observatory, Department of Astronomy and Theoretical Physics, Box 43, SE-221 00 Lund, Sweden\\
$^3$INAF - Osservatorio Astrofisico di Arcetri, Largo E. Fermi 5, 50125, Florence, Italy\\
$^4$Max-Planck Institut f\"{u}r Astronomie, K\"{o}nigstuhl 17, 69117 Heidelberg, Germany\\
$^5$Department of Physics and Astronomy, Uppsala University, Box 516, SE-751 20 Uppsala, Sweden\\
$^6$Astrophysics Group, Keele University, Keele, Staffordshire ST5 5BG, United Kingdom\\
$^7$INAF - Padova Observatory, Vicolo dell'Osservatorio 5, 35122 Padova, Italy\\
$^8$INAF - Osservatorio Astronomico di Bologna, via Ranzani 1, 40127, Bologna, Italy\\
$^9$INAF - Osservatorio Astronomico di Palermo, Piazza del Parlamento 1, 90134, Palermo, Italy\\
$^{10}$GEPI, Observatoire de Paris, CNRS, Universit\'e Paris Diderot, 5 Place Jules Janssen, 92190 Meudon, France\\
$^{11}$Dipartimento di Fisica e Astronomia, Sezione Astrofisica, Universit\'{a} di Catania, via S. Sofia 78, 95123, Catania, Italy\\
$^{12}$ASI Science Data Center, Via del Politecnico SNC, 00133 Roma, Italy\\
$^{13}$Laboratoire Lagrange (UMR7293), Universit\'e de Nice Sophia Antipolis, CNRS,Observatoire de la C\^ote d'Azur, CS 34229,F-06304 \\Nice cedex 4, France\\
$^{14}$Department for Astrophysics, Nicolaus Copernicus Astronomical Center, ul. Rabia\'{n}ska 8, 87-100 Toru\'{n}, Poland\\
$^{15}$European Southern Observatory, Alonso de Cordova 3107 Vitacura, Santiago de Chile, Chile\\
$^{16}$Instituto de Astrof\'{i}sica de Andaluc\'{i}a-CSIC, Apdo. 3004, 18080 Granada, Spain\\
$^{17}$Universit\`a di Bologna, Dipartimento di Fisica e Astronomia, viale Berti Pichat 6/2, 40127 Bologna, Italy\\
$^{18}$INAF - Osservatorio Astrofisico di Catania, via S. Sofia 78, 95123, Catania, Italy\\
$^{19}$Astrophysics Research Institute, Liverpool John Moores University, 146 Brownlow Hill, Liverpool L3 5RF, United Kingdom\\
$^{20}$Departamento de Ciencias Fisicas, Universidad Andres Bello, Republica 220, Santiago, Chile\\
$^{21}$Millennium Institute of Astrophysics, Chile\\
$^{22}$Pontificia Universidad Cat\'{o}lica de Chile, Av. Vicu\~{n}a Mackenna 4860, 782-0436 Macul, Santiago, Chile\\
$^{23}$Instituto de Astrof\'isica e Ci\^encias do Espa\c{c}o, Universidade do Porto, CAUP, Rua das Estrelas, 4150-762 Porto, Portugal\\
$^{24}$Institute of Theoretical Physics and Astronomy, Vilnius University, A. Gostauto 12, LT-01108 Vilnius\\
$^{25}$Faculty of Mathematics and Physics, University of Ljubljana, Jadranska 19, 1000, Ljubljana, Slovenia\\
$^{26}$School of Physics, University of New South Wales, Sydney NSW 2052, Australia\\
$^{27}$Stellar Astrophysics Centre, Department of Physics and Astronomy, Aarhus University, Ny Munkegade 120, DK-8000 Aarhus C, Denmark\\
$^{28}$School of Physics and Astronomy, University of Birmingham, Edgbaston, Birmingham B15 2TT, United Kingdom\\
$^{29}$Leibniz-Institut f\"ur Astrophysik Potsdam (AIP), An der Sternwarte 16, 14482 Potsdam, Germany\\
$^{30}$Department of Physics and Astronomy G. Galilei,  University of Padova, Vicolo dell'Osservatorio 3, I-35122 Padova, Italy\\
$^{31}$Institut d'Astrophysique et de G\'eophysique, Universit\'e de Li\`ege, Quartier Agora, B\^at. B5c, All\'ee du 6 Ao\^ut, 19c 4000 Li\`ege, Belgium\\
\label{lastpage}

\end{document}